\documentclass[draft]{agujournal2019}
\usepackage{url} 
\usepackage{lineno}
\usepackage[inline]{trackchanges} 
\usepackage{soul}
\draftfalse

\journalname{Geochemistry, Geophysics, Geosystems}

\usepackage{amsmath}
\newcommand{\cotwo}{CO$_2$ }
\newcommand{\OLR}{\textrm{OLR}}
\newcommand{\Tsnow}{T_{\textrm{snow}}}

\begin{document}

%
%


\title{Simple Stochastic Modeling of Snowball Probability Throughout Earth History}

%
%




\authors{Mark Baum, Minmin Fu\affil{1,2}}

\affiliation{1}{Department of Earth and Planetary Sciences, Harvard University}
\affiliation{2}{Department of Earth and Planetary Sciences, Yale University}





\correspondingauthor{Mark Baum}{markmbaum@protonmail.com}




\begin{keypoints}
\item We explore models of randomness in the long-term climate and carbon cycle
\item Symmetric noise in the \cotwo concentration is instructive, but incompatible with the snowball record
\item Modeling noise in the \cotwo outgassing rate, instead, is more flexible and can be qualitatively reconciled with the snowballs
\end{keypoints}

%
%

%
%


\begin{abstract}
Over its multibillion-year history, Earth has exhibited a wide range of climates. Its history ranges from snowball episodes where the surface was mostly or entirely covered by ice to periods much warmer than today, where the cryosphere was virtually absent. Our understanding of greenhouse gas evolution over this long history, specifically carbon dioxide, is mainly informed by deterministic models. However, the complexity of the carbon cycle and its uncertainty over time motivates the study of non-deterministic models, where key elements of the cycle are represented by inherently stochastic processes. By doing so, we can learn what models of variability are compatible with Earth's climate record instead of how exactly this variability is produced. In this study, we address why there were snowballs in the Proterozoic, but not the Phanerozoic by discussing two simple stochastic models of long-term carbon-cycle variability. The first, which is the most simple and represents \cotwo concentration directly as a stochastic process, is instructive and perhaps intuitive, but is incompatible with the absence of snowballs in the Phanerozoic. The second, which separates carbon source from sink and represents \cotwo outgassing as a stochastic process instead of concentration, is more flexible. When outgassing fluctuates over longer periods, as opposed to brief and rapid excursions from a mean state, this model is more compatible with the snowball record, showing only modest increases in the probability of snowball events over Earth history. The contrast between these models illustrates what kind of variability may have characterized the long-term carbon cycle.
\end{abstract}

\section*{Plain Language Summary}
The Earth's long-term climate record is complicated and variable. At some points, our planet has been extremely cold and almost entirely covered in ice (snowball periods). At others, the climate was much warmer than today and the surface was ice-free. These changes may have many overlapping causes and are quite complex. In light of this, we use simple models to understand how randomness in the carbon cycle affects the long-term climate. In one model, the atmospheric carbon dioxide concentration is treated as a randomly fluctuating quantity. We can learn from this model, but it is at odds with the timing of Earth's snowball periods. In a second model, the \cotwo \emph{outgassing rate} is represented by a randomly fluctuating quantity instead of the concentration itself, and is more consistent with the geological record. These models offer some qualitative lessons about the kind of randomness that is compatible with Earth's snowball episodes.

%
%

\section{Introduction}

Earth's climate has varied dramatically over the past 2.5 billion years, ranging from ``hothouse" climates such as the Eocene (55.8 -- 33.9 Ma) \cite{Berner-1990:atmospheric, Greenwood-Wing-1995:eocene} to ``Snowball Earth" episodes, when virtually all of the Earth's surface was encased in ice \cite{hoffman_snowball_2002, pierrehumbert_climate_2011}. Geological evidence points to at least three snowball episodes, with the first event occurring at the beginning of the Paleoproterozoic era, at approximately 2.5~Ga \cite{evans_low-latitude_1997, kirschvink_paleoproterozoic_2000}. The next two occurred during the Cryogenian period (720 -- 635 Ma), with inception times about 50~Myr apart \cite{Hoffman-Kaufman-Halverson-et-al-1998:neoproterozoic, Rooney-Strauss-Brandon-et-al-2015:cryogenian, Prave-Condon-Hoffmann-et-al-2016:duration}. 

What caused snowball climates at some points, but warm climates at others? Over geologic timescales, the atmospheric carbon dioxide (\cotwo) concentration, and therefore global temperature, is set by the balance between sources (volcanic outgassing) and sinks (silicate weathering) \cite{Urey-1952:early, Walker-Hays-Kasting-1981:negative, Marshall-Walker-Kuhn-1988:long, Berner-Lasaga-1989:modeling}. Silicate weathering and marine carbonate burial sequester \cotwo from the ocean and atmosphere until it is recycled back by the subduction of oceanic crust \cite{Berner-Lasaga-Garrels-1983:carbonate}. Indeed, variations in volcanic outgassing and availability of weatherable materials have both been suggested to be dominant controls on Earth's long term climate evolution \cite<e.g.,>{McKenzie-Horton-Loomis-et-al-2016:continental,macdonald_arc-continent_2019}.

Earth's temperature and \cotwo concentration are not well known deep into the planet's history. Although the Sun's luminosity has slowly increased since the birth of the solar system, there is no evidence for secular warming of the climate \cite{feulner2012faint}. To compensate for the increase in absorbed solar radiation, the mean \cotwo concentration has very likely decreased over time \cite{Halevy-Bachan-2017:geologic,Krissansen-Totton-Arney-Catling-2018:constraining}. This decrease has not been smooth, but rather accompanied by significant variability, the cause of which may be multifactorial and is still an active area of research \cite{franks2014new, montanez2016climate, lenardic2016climate, macdonald_arc-continent_2019, park2021evaluating, baum2022sensitive}.

Past studies have generally attempted to understand variability in atmospheric \cotwo concentration in a deterministic fashion, by investigating specific components of the carbon cycle and their potential effect on climate. For example, we can attempt to constrain the independent effects of continental configuration, topography, biochemistry, ocean circulation, and a wide range of other processes. Here we take a different approach, investigating the effects of explicitly random or stochastic processes in the long term carbon cycle. This approach views the climate system as an inherently stochastic, complex system and attempts to understand what models of randomness agree with our sparse and incomplete record of Earth's climate on the longest timescales. Specifically, we anchor our modeling and discussion to the aforementioned Snowball Earth episodes, as they represent very important excursions from the mean state. Our conception of long-term climate variability must be compatible with the observed timing of snowball events.

That is not to say that stochastic models are absent from the climate literature. There is a long history of stochastic climate modeling touching virtually all components of the Earth system \cite{hasselmann_stochastic_1976, imkeller2001stochastic, imkeller2002conceptual, Farrell-Abbot-2012:mechanism, franzke2015stochastic}. Some recent studies have used stochastic methods to model Earth's carbon cycle \cite<e.g.,>{Zeebe-Westerhold-Littler-et-al-2017:orbital,wordsworth_how_2021,Arnscheidt-Rothman-2021:asymmetry}. In this paper, we take a similar approach by setting the complexity of the three-dimensional, chaotic climate system aside to examine the consequences of two simple stochastic models of the carbon cycle over the past $\sim$2~Gyr. We do not suggest that our modeling is a precise representation of Earth's long-term carbon record, but use it to understand what kinds of variability (what models) are compatible with the snowball record.

At the beginning of Section \ref{sec:modeling}, we explain the primary physical equations and assumptions governing both of these models. In Section \ref{sec:stochastic_co2}, we interrogate the stochastic model recently published by \citeA{wordsworth_how_2021} and discuss its compatibility with Earth's record of snowball periods. In \ref{sec:stochastic_outgassing} we describe an alternative model with a deterministic weathering sink but stochastic \cotwo outgassing. In Section \ref{sec:simulations} we explain how we solve the model equations numerically and produce a large ensemble of these solutions. In Section \ref{sec:resdis} we present the results of these simulations, compare with the model of Section \ref{sec:stochastic_co2}, and discuss compatibility with Earth's snowball record. Finally, in Section \ref{sec:conclusion}, we offer some concluding remarks and discuss possible avenues of future research.

\section{Simple Stochastic Climate Modeling}
\label{sec:modeling}

We use a simple, zero-dimensional model of stochastic, long-term climate variability representing the long-term evolution Earth's carbon cycle. The time-evolution of Earth's temperature is governed by the balance between outgoing longwave radiation (OLR) and absorbed solar radiation,
\begin{linenomath*}
\begin{equation}
C \dot{T} = \frac{F(t)}{4}\left[ 1 - \alpha \right] - \OLR(T,f) \, ,
\label{eq:govern}
\end{equation}
\end{linenomath*}
where $C$ is the system's heat capacity, $T$ is the global mean surface temperature, $t$ is time, $F$ is the time-dependent incident solar flux, $\alpha$ is the planetary albedo, OLR is the temperature-dependent and CO$_2$-dependent outgoing longwave radiation, and $f$ is the atmospheric \cotwo concentration in ppm. We set the albedo to a constant 30~\%.

In reality, due to the response of clouds and ice to surface temperature, albedo is a complex function of temperature. However, the goal of this model is to simulate the likelihood of different temperatures over time, not represent the full-complexity of the ice-albedo feedback. Instead, we assume a hypothetical snowball threshold temperature of $\Tsnow=280$~K \cite{pierrehumbert_climate_2011} where the ice-albedo feedback runs away. The precise value of this threshold does not affect our conclusions.

The solar flux increases over time according to the standard approximation \cite{gough_solar_1981},
\begin{linenomath*}
\begin{equation}
F(t) = F_0 \left[ 1 + \frac{2}{5}\left( 1 - t/t_0 \right) \right] ^{-1} \, ,
\end{equation}
\end{linenomath*}
where $F_0$ is the modern value of 1366~W/m$^2$ and $t_0$ is 4.5~Gyr.

The heat capacity of the ocean-atmosphere system is such that the temperature approaches equilibrium over $\sim$10$^3$~yr, much more rapidly than the 10$^6$-10$^9$~yr timescales of interest in our simulations. Accordingly, we assume instantaneous temperature equilibrium and set $\dot{T}$ to zero. We also linearize OLR around Earth's preindustrial surface temperature and \cotwo concentration of $T_e=288$~K and $f_0=285$~ppm, respectively.
\begin{linenomath*}
\begin{equation}
\OLR = \OLR_0 + a \left( T - T_e \right) - b \log \left( f/f_0 \right) \, ,
\label{eq:linearization}
\end{equation}
\end{linenomath*}
where $a=2$~W/m$^2$/K \cite{abbot2016analytical} and $b=5.35$~W/m$^2$ \cite{myhre1998new}. To balance Equation \ref{eq:govern} at preindustrial conditions and $t=4.5$~Gyr, OLR$_0$ must be equal to $F_0(1 - A)/4$. These assumptions, plugged into Equation \ref{eq:govern}, yield
\begin{linenomath*}
\begin{equation}
0 = \frac{F(t)}{4}\left[ 1 - \alpha \right] - \OLR_0 - a \left( T - T_e \right) + b \log \left( f/f_0 \right)
\end{equation}
\end{linenomath*}
which can be solved directly for temperature as a function of time and \cotwo concentration,
\begin{linenomath*}
\begin{equation}
T(t,f) = T_e + \frac{1}{a}\left[ \frac{F(t)}{4} \left( 1 - \alpha \right) - \OLR_0 + b\log\left( f/f_0 \right) \right] \, .
\label{eq:T}
\end{equation}
\end{linenomath*}

This is the governing equation of all our subsequent modeling. We note that the linearization in Equation \ref{eq:linearization} is a fairly good approximation when $T$ is within 210--310~K \cite{Koll-Cronin-2018:earth}.

\subsection{Stochastic \cotwo Concentration}
\label{sec:stochastic_co2}

Equation \ref{eq:T} defines temperature in terms of time $t$ (capturing the brightening sun) and the carbon dioxide concentration $f$. To introduce randomness into this system, one approach is to model $f$ as a stochastic process. We analyze and discuss this approach here, along with its primary assumptions and implications, before continuing with the development of our new model equations.

Recently, \citeA{wordsworth_how_2021} used an Ornstein-Uhlenbeck process \cite{jacobs_stochastic_2010, dobrow_introduction_2016} to represent randomness in $f$. This approach models $f$ with Gaussian noise that relaxes to a prescribed value $\chi$. They used a nondimensional representation, $y = f/f_0$, and the model equation is
\begin{linenomath*}
\begin{equation}
dy = \frac{1}{\tau}\left[ y - \frac{\chi(t)}{f_0} \right] dt + \sigma dB \, ,
\label{eq:df}
\end{equation}
\end{linenomath*}
where $\tau$ is the relaxation timescale, $\chi$ is the time-dependent value of $f$ that would achieve perfect temperature equilibrium (at $T_e$), $\sigma$ scales the noise, and $dB$ is a Wiener process (Gaussian or Brownian noise). In this case, $\tau$ is analogous to the timescale required for chemical weathering to restore temperature equilibrium.

\begin{figure}
\centering
\includegraphics[width=0.9\textwidth]{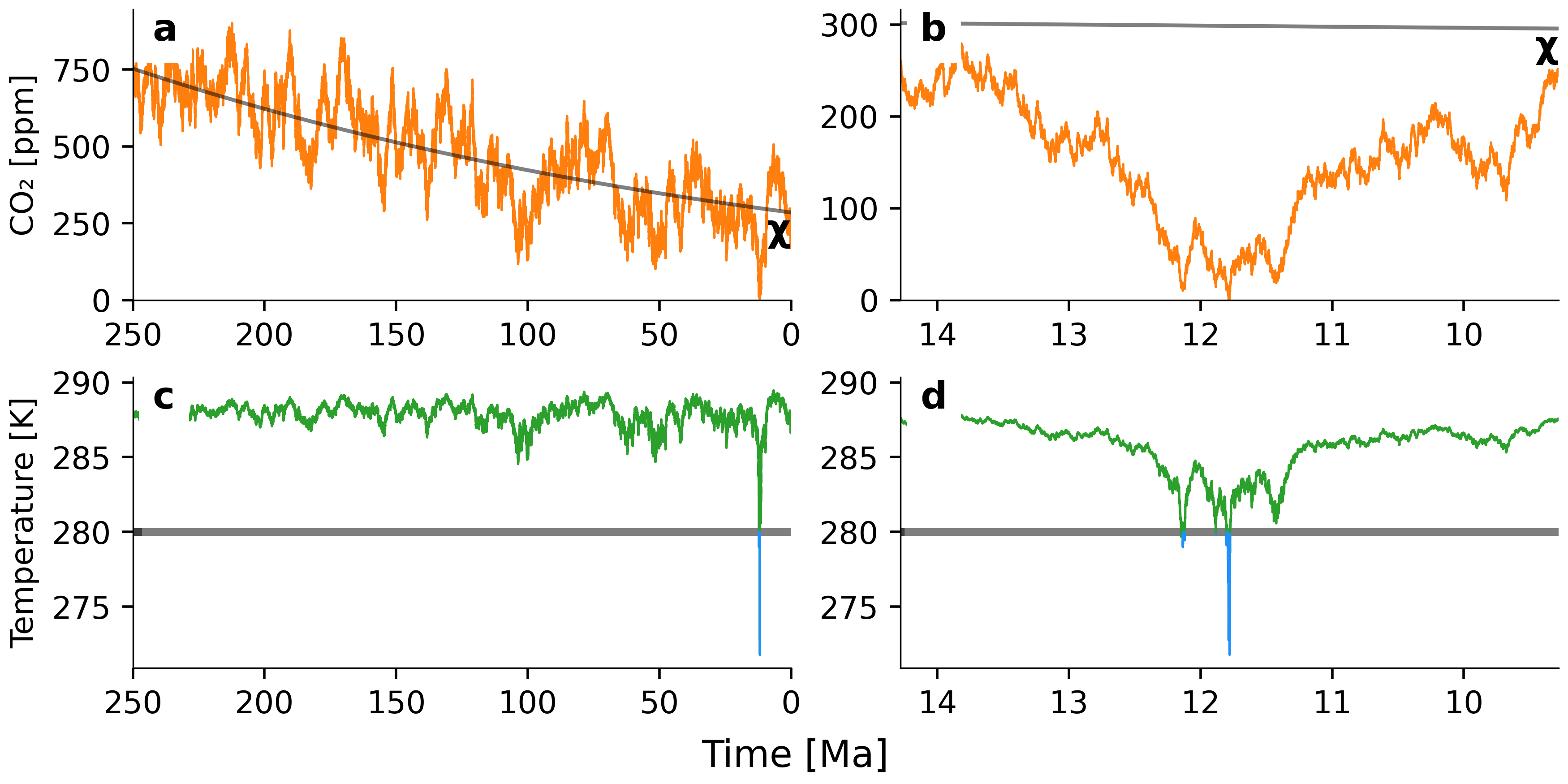}
\caption{A single realization of the stochastic \cotwo concentration model described by Equations \ref{eq:T} and \ref{eq:df}, over 250~Myr, with ${\tau=3}$~Myr and ${\sigma=10}$. Panels (a) and (b) show the \cotwo concentration over time, with the equilibrium value $\chi$ drawn in gray. Panel (a) shows the entire time series and panel (b) shows the same simulation, but zoomed into the $\sim$5~Myr surrounding the temperature minimum. Panels (c) and (d) show temperature evolution for the same simulation, again zoomed in for panel (d), with temperatures below 280~K highlighted in blue. Panels (b) and (d), in particular, highlight the exquisite temperature sensitivity when \cotwo concentrations are low. A relatively small change in $f$ near the minimum between 13 and 11~Ma produces a very dramatic drop in temperature. Note that sharp negative excursions are a consequence of the logarithmic dependence of temperature on CO$_2$, and not an ice-albedo feedback.}
\label{fig:statT}
\end{figure}

In their model, randomness occurs \emph{directly} in the \cotwo concentration and $f$ is the sole prognostic variable. The stochastic differential equation above is integrated numerically over 10$^8$--10$^9$~yr and the resulting sequence of discrete $f$ values determines the simulated temperature history via Equation \ref{eq:T}. As time proceeds, $f$ varies randomly but is restored to equilibrium over a timescale of $\tau\approx3$~Ma, which is short compared to the integration timescale. Figure \ref{fig:statT} shows one example simulation using this equation, where $\sigma$ was chosen for a reasonable chance that the temperature would reach $\Tsnow$.

Relaxation of $f$ toward $\chi$ happens considerably faster than the slow, insolation driven drift in $\chi$ itself. As a result, over many random realizations of this model, the \cotwo concentration is approximately normally distributed around $\chi$ at all points in time.
\begin{linenomath*}
\begin{equation}
f \sim \mathcal{N}(\chi(t), \sigma^2) \, ,
\end{equation}
\end{linenomath*}
where $\mathcal{N}$ is the normal distribution with mean of $\chi(t)$ and variance of $\sigma^2$. It is relatively straightforward to examine the resulting temperature probabilities over time, using Equation \ref{eq:T}.

\begin{figure}
\centering
\includegraphics[width=0.9\textwidth]{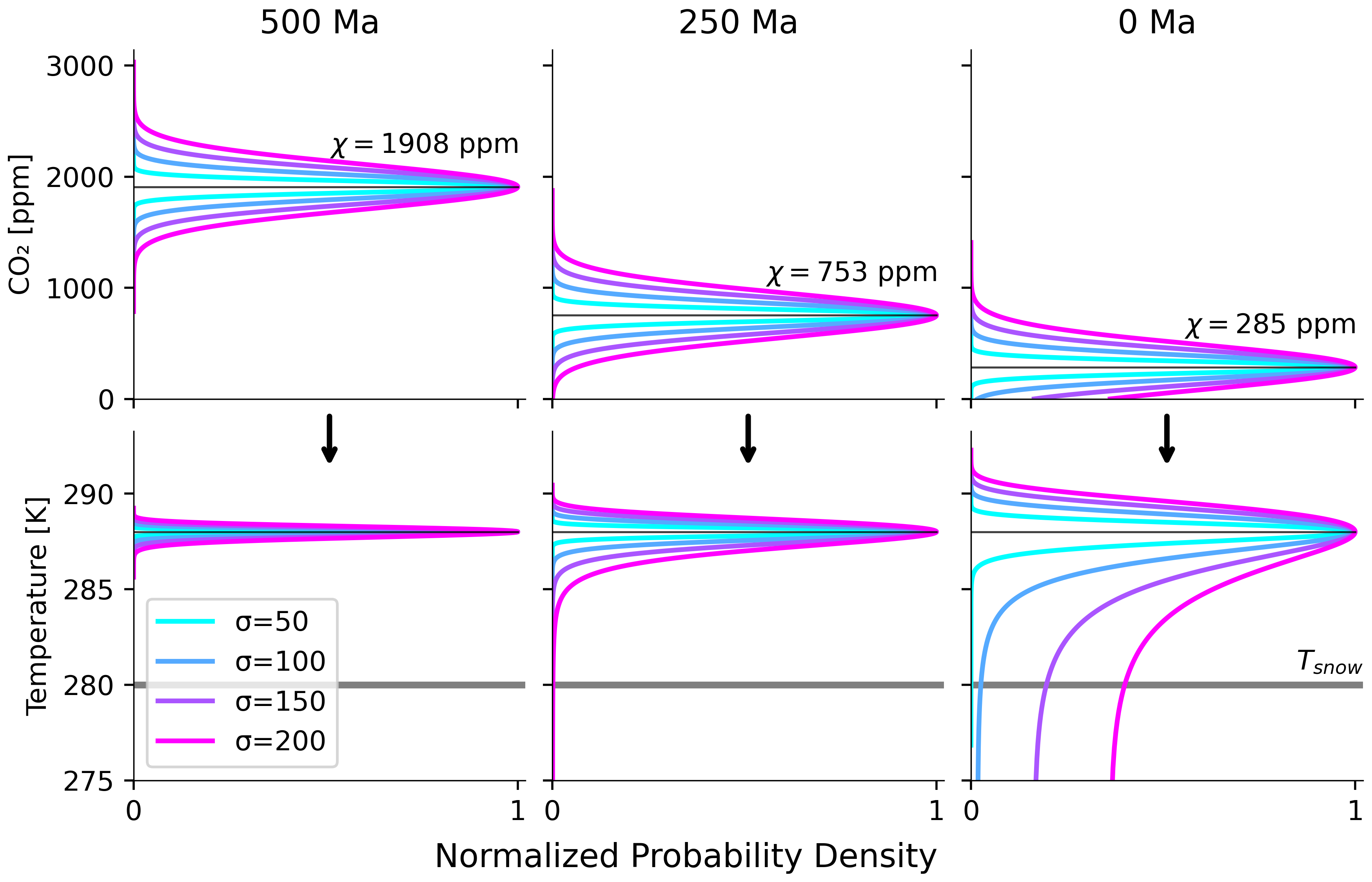}
\caption{In the top row, normal distributions in the atmospheric \cotwo concentration ($f$), for several $\sigma$ values, at three different snapshots in time. The $\chi$ values are the distribution means required to produce a mean temperature of ${T_e = 288}$~K using Equation \ref{eq:T}. The bottom row shows the exact same distributions, but plotted in temperature space instead. These plots show how, even though the mean temperature does not change (by design), the space of probable temperatures widens over time and cold temperatures become strongly favored. The logarithmic relationship between $T$ and $f$ produces this effect. The hypothetical snowball temperature of 280~K becomes dramatically more likely at later times.}
\label{fig:statdist}
\end{figure}

The top row of Figure \ref{fig:statdist} shows distributions in the \cotwo concentration $f$ at three different times over the most recent 500~Myr. Earlier in time, the sun is fainter so a higher $\chi$ value is required for a mean temperature of 288~K. At later times, the brighter sun results in lower $\chi$ values, shifting the distribution down. The bottom row of this figure shows the same distributions, but plotted against the corresponding temperatures. Because of the logarithmic relationship between $T$ and $f$, the probable temperature space widens as time proceeds and cold temperatures become significantly more likely than warm ones. This happens because, for a single value of $\sigma$, the variation in $f$ represents a larger fractional change when the mean is lower. In this model, snowball temperatures are dramatically more likely at later times. This phenomena was also recently discussed by \citeA{Graham-2021:high}.

\begin{figure}
\centering
\includegraphics[width=0.9\textwidth]{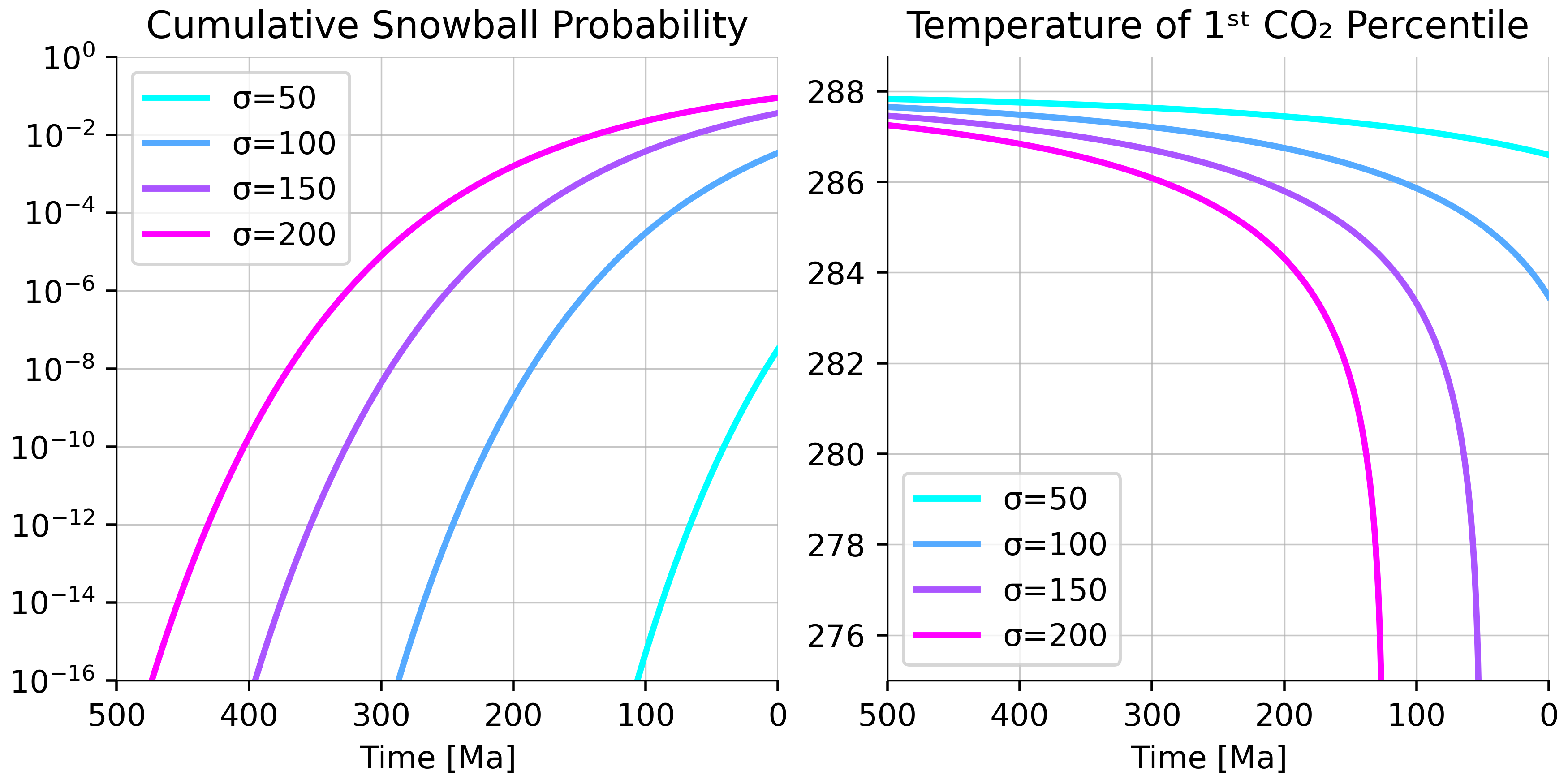}
\caption{On the left, the cumulative probability of snowball temperatures for different values of $\sigma$ over time. Each line represents the probability that random variation in $f$, as defined by Equation \ref{eq:df}, would produce $T \leq 280$~K. As expected from Figure \ref{fig:statdist}, the snowball probability plummets at earlier times. On the right, the temperatures corresponding to the first percentile of the $f$ distributions. Although the first percentile is an arbitrary metric, it illustrates the very strong bias against cold temperatures at earlier times.}
\label{fig:statprob}
\end{figure}

Figure \ref{fig:statprob} shows exactly how much more likely cold excursions become at later times in this model: orders of magnitude. The left panel shows the cumulative probability of the \cotwo concentration required to achieve a hypothetical snowball temperature of 280~K, for several values of $\sigma$, over time. Note the logarithmic vertical axis. Beyond the most recent periods of time, a snowball is effectively impossible. The right panel shows the temperature corresponding to the first percentile of the $f$ distributions over time, showing a precipitous drop in this temperature as time proceeds.

To summarize, this model represents randomness in the climate system by making the \cotwo concentration approximately equivalent to a normally distributed random variable with a mean value that drifts over time to maintain stable temperature. Outside the most recent period, stochastically induced cold temperatures are extremely low probability---effectively prohibited. For higher values of $\sigma$, as \citeA{wordsworth_how_2021} showed, the same transition occurs as time proceeds, but it simply occurs at earlier times.

However, it is worth considering whether randomness in the climate system is well represented by Gaussian noise in the \cotwo concentration directly. First, the principal results of this model strongly contradict Earth's geologic record, which indicates snowball episodes occurred roughly 2.5~Ga and 700~Ma. To be fair, \citeA{wordsworth_how_2021} focuses mainly on application to exoplanets, but in that case assuming Earth-like weathering sinks and Sun-like stellar properties considerably limits the model's applicability.
 
Second, it is questionable to assume that \cotwo concentrations are well represented by symmetric noise. To first order, atmospheric \cotwo is controlled over long periods by volcanic outgassing (source) and chemical weathering (sink). It is intuitive to assume that volcanic outgassing is noisy, as it is comprised of eruptions of various magnitudes and timescales. This justifies a noisy, non-negative, \emph{source} of carbon dioxide. While the weatherability of Earth's crust has probably varied considerably over Earth history \cite{macdonald_arc-continent_2019}, these fluctuations occur on longer timescales and are unlikely to resemble the character of noisy volcanic outgassing.

This model yields conceptual insights. It demonstrates the basic point that any change in atmospheric \cotwo concentration produces a larger temperature response when the initial concentration is lower. This response is strongly skewed toward cold temperatures because of the temperature's logarithmic dependence on CO$_2$. Figure \ref{fig:statdist} demonstrates these effects. However, this model contradicts Earth's snowball record and may be of limited relevance to exoplanets because of its strong assumptions about Earth-like weathering behavior and boundary conditions. It also makes no explicit distinction between weathering and outgassing, limiting its physical interpretability.

\subsection{Stochastic \cotwo Outgassing}
\label{sec:stochastic_outgassing}

We develop a model with a more physically consistent representation of long-term \cotwo outgassing and removal. In our model, the temperature is still driven by \cotwo concentration and the secular brightening of the sun defined by Equation \ref{eq:T}. However, in this case the \cotwo concentration is not a stochastic variable itself, but is governed by the balance between deterministic weathering (sink) and stochastic volcanic outgassing (source).
\begin{linenomath*}
\begin{equation}
\dot{C}(t) = V(t) - W(C) \, ,
\end{equation}
\end{linenomath*}
where $C$ is the total ocean atmosphere-ocean reservoir of \cotwo, $V$ is the volcanic \cotwo outgassing rate, and $W$ is the rate of \cotwo sequestration via silicate weathering. Weathering is prescribed by the traditional WHAK formulation \cite{Walker-Hays-Kasting-1981:negative}, where the weathering rate is exponentially dependent on temperature changes,
\begin{linenomath*}
\begin{equation}
W(T) = k \exp\left( \frac{T - T_e}{T_s} \right) \, .
\label{eq:whak}
\end{equation}
\end{linenomath*}
Here, $k$ is a calibration constant and $T_s$ is the exponential scaling factor. Note that we exclude direct dependence on carbon dioxide concentration because, with it, weathering can no longer be calibrated to the same equilibrium temperature at different solar forcing \cite{pierrehumbert_principles_2010}.

To evaluate the weathering equation, we convert the total carbon $C$ to temperature $T$ by first converting to concentration $f$ in ppm. To do this, we use the simple ocean-atmosphere partitioning model of \citeA{mills_timing_2011},
\begin{linenomath*}
\begin{equation}
p = 0.78\frac{C}{C + h} \, ,
\end{equation}
\end{linenomath*}
where $p$ is the \cotwo partial pressure and $h$ is a constant 2.33$\times 10^{8}$ Tmole \cotwo. From there, we assume a background partial pressure of approximately 1 bar to compute the \cotwo concentration $f$, which is used to evaluate Equation \ref{eq:T}. Finally, the resulting temperature is used to compute the weathering and sequestration rate.

The outgassing rate $V$ fluctuates stochastically and is defined by an Ornstein-Uhlenbech process,
\begin{linenomath*}
\begin{equation}
dV = \frac{1}{\tau} \left( \mu - V \right) dt + \sigma dB \, ,
\end{equation}
\end{linenomath*}
where $\tau$ is a prescribed relaxation time scale, $\mu$ is a prescribed mean outgassing rate, $\sigma$ scales the noise term, and $dB$ is a Weiner process (Gaussian or Brownian noise). Note that this equation and its parameters are distinct from Equation \ref{eq:df}. Here, $\tau$ no longer represents a weathering time scale because this stochastic process represents outgassing, not \cotwo concentration directly.

All together, our model consists of two differential equations. The first is a straightforward definition of how total ocean-atmosphere carbon changes in response to the balance of weathering and outgassing. The second defines stochastic variation in the \cotwo outgassing rate.
\begin{linenomath*}
\begin{align}
\dot{C} &= V(t) - W(C)  \label{eq:C} \\[1ex]
\dot{V} &= \frac{1}{\tau} \left( \mu - V \right) + \sigma B(t)  \label{eq:V}
\end{align}
\end{linenomath*}

The difference between this model and the one discussed in Section \ref{sec:stochastic_co2} is that, here, neither the carbon reservoir $C$ nor the \cotwo concentration $f$ are \emph{themselves} stochastic processes. The carbon sink and source terms are separate. Weathering is deterministic and depends implicitly on temperature, as defined in Equation \ref{eq:whak}. The outgassing \emph{rate} $V$ is stochastic and modeled with an Ornstein-Uhlenbeck process. In contrast to the model in Section \ref{sec:stochastic_co2}, these equations explicitly represent imbalances in the carbon system that drive long-term temperature change.

\section{Simulations}
\label{sec:simulations}

We integrate the system of Equations \ref{eq:C} and \ref{eq:V} using the simple Euler-Maruyama method over a range of different parameters and analyze the statistics of the ensemble. We vary $\tau$ and $\sigma$ over discrete values between 10$^5$--10$^9$~yr and 10$^{-5}$--10$^{-2}$~Tmole/yr, respectively, performing many simulations with each available combination. We execute simulations in two stages. First, we perform a relatively small number of simulations over each combination of $\tau$ and $\sigma$. Second, we identify which combinations yield simulations with appreciable, but not unrealistically high, temperature variability and perform a much larger number of simulations using those combinations. This is a simple strategy to avoid wasting computation on inconsequential $(\tau,\sigma)$ pairs. All told, the ensemble is comprised of 27,551,040 unique simulations.

Each simulation begins at $t=2$~Ga, when the equilibrium \cotwo concentration is slightly above 10$^5$ ppm, and ends at 0~Ga (present day). Each simulation is also taken through a spin-up period of 0.5~Gyr to avoid starting them at the exact equilibrium temperature, then integrated using 1 million time steps. To prevent evaluating the logarithm of a non-positive number, $V$ is restricted to be above zero and $C$ is restricted to be above a very small fraction of its equilibrium value at modern insolation. These restrictions have virtually no impact on our conclusions and only serve to prevent computation errors with parameter combinations that produce unrealistically high variability. For each simulation, we store the values of $C$, $V$, $T$, and $W$ at 17 equally spaced points in time. We also record the maximum and minimum value of each quantity along with the precise time of the temperature extrema. Table \ref{tab:params} consolidates all of the model parameters along with their values and units.

\begin{table}
\caption{Static model parameters with values and units}
\centering
\begin{tabular}{l l l l}
\hline
Parameter & Description & Value/Range & Units \\
\hline
$f_0$ & preindustrial \cotwo concentration & 285 & ppm \\
$T_e$ & preindustrial temperature & 288 & K \\
$\alpha$ & planetary albedo & 0.3 & dimensionless \\
OLR$_0$ & equilibrium outgoing longwave radiation & 239.05 & W/m$^2$ \\
$h$ & ocean-atmosphere partitioning constant & 2.33$\times$10$^{8}$ & Tmole \\
$\mu$ & mean volcanic outgassing rate & 7 & Tmole/yr \\
$t_1$ & simulation start time & 2.5 & Ga \\
$t_2$ & simulation end time & 0 & Ga \\
$t_0$ & temporal scaling of insolation function & 4.5 & Ga \\
$t_{\textrm{spinup}}$ & model spinup duration & 0.5 & Ga \\
$T_s$ & weathering temperature scaling & 11.1 & K \\
$\Tsnow$ & snowball threshold temperature & 280 & K \\
$k$ & weathering calibration constant & 7 & Tmole/yr \\
$a$ & OLR temperature scaling & 2 & W/m$^2$/K \\
$b$ & OLR $\log f$ scaling & 5 & W/m$^2$ \\
$F_0$ & modern solar insolation & 1366 & W/m$^2$ \\
$\tau$ & outgassing relaxation timescale & 10$^5$ - 10$^9$ & yr \\
$\sigma$ & outgassing deviation/variability & 10$^{-5}$ - 10$^{-2}$ & Tmole/yr
\end{tabular}
\label{tab:params}
\end{table}

\section{Results and Discussion}
\label{sec:resdis}

\begin{figure}
\centering
\includegraphics[width=0.8\textwidth]{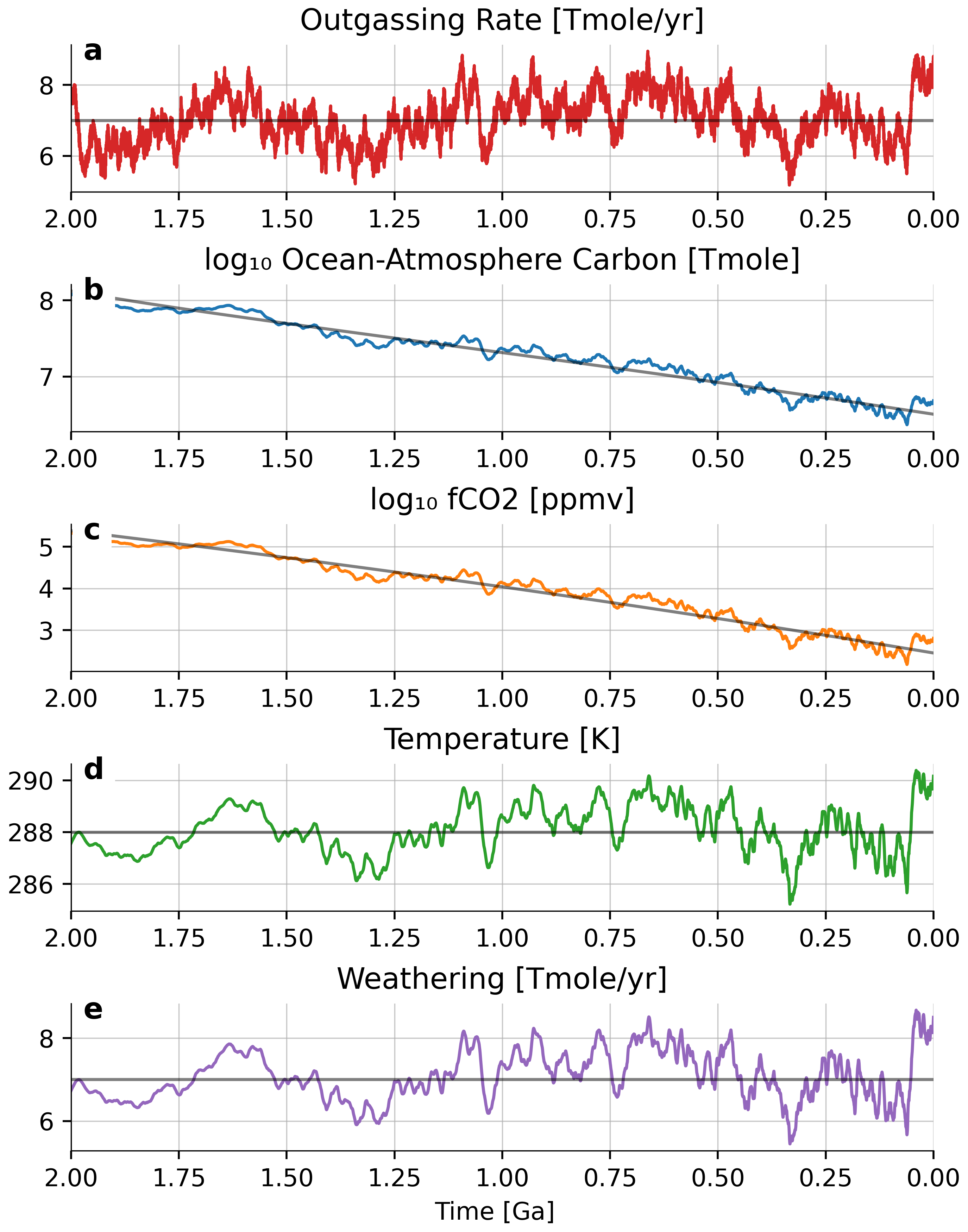}
\caption{An example simulation of the system defined by Equations \ref{eq:C} and \ref{eq:V}. In panel (a), the stochastic outgassing rate $V$. In panel (b), the total atmosphere-ocean \cotwo reservoir $C$. These are the prognostic model variables. Each of the remaining quantities is panels (c-e) is derived from $C$. The gray line in each panel shows the equilibrium value or the mean value appropriate for each quantity. For the outgassing and weathering in panels (a) and (e) it shows the mean value of 7~Tmole/yr. In panels (b) and (c) it shows the value required for the equilibrium temperature of $T_0=288$~K, over time. Finally, in panel (d), it shows the equilibrium temperature of 288~K.}
\label{fig:exsim}
\end{figure}

For orientation, Figure \ref{fig:exsim} shows the results of a single simulation with $\tau=30\times 10^6$~yr and $\sigma=2\times 10^{-4}$~Tmole/yr. The volcanic outgassing rate, which is a realization of the Ornstein-Uhlenbeck process defined in Equation \ref{eq:V}, is shown on the top in red. The total atmosphere-ocean carbon stock is shown on a log scale just below in blue. These are the prognostic variables of the model. The \cotwo concentration, temperature, and weathering rate are each derived from $C$ as described in Section \ref{sec:modeling}.

\begin{figure}
\centering
\includegraphics[width=0.9\textwidth]{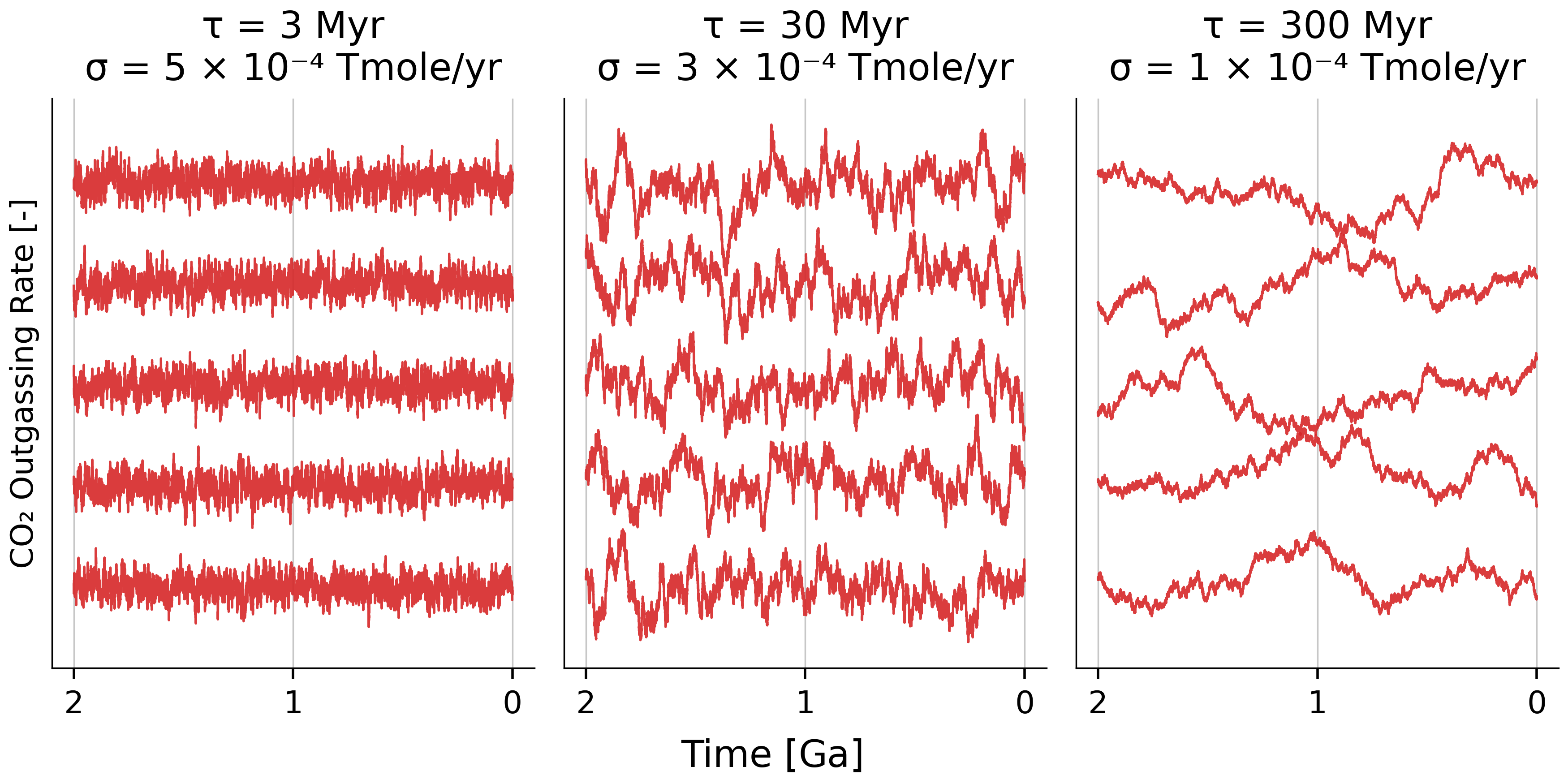}
\caption{Realizations of the Ornstein-Uhlenbeck process defined by Equation \ref{eq:V} for three exemplary pairs of $\tau$ and $\sigma$. Five independent realizations are stacked vertically on top of each other.}
\label{fig:outexamples}
\end{figure}

Figure \ref{fig:outexamples} shows more realizations of the stochastic outgassing rate for different combinations of $\tau$ and $\sigma$, stacked on top of each other. When $\tau$ is small, the outgassing rate quickly relaxes to the mean and the value of $\sigma$ must be relatively high to produce random fluctuations of appreciable magnitude. In this regime, changes in the outgassing rate are dominated by short-term, rapid excursions. In an intermediate regime, with $\tau=30$~Myr, longer-term fluctuations begin to appear because the relaxation time scale is longer, but rapid excursions are still evident. Finally, with $\tau=300$~Myr, the outgassing rate is dominated by longer period drift with occasional and smaller magnitude short-term shocks.

\begin{figure}
\centering
\includegraphics[width=0.9\textwidth]{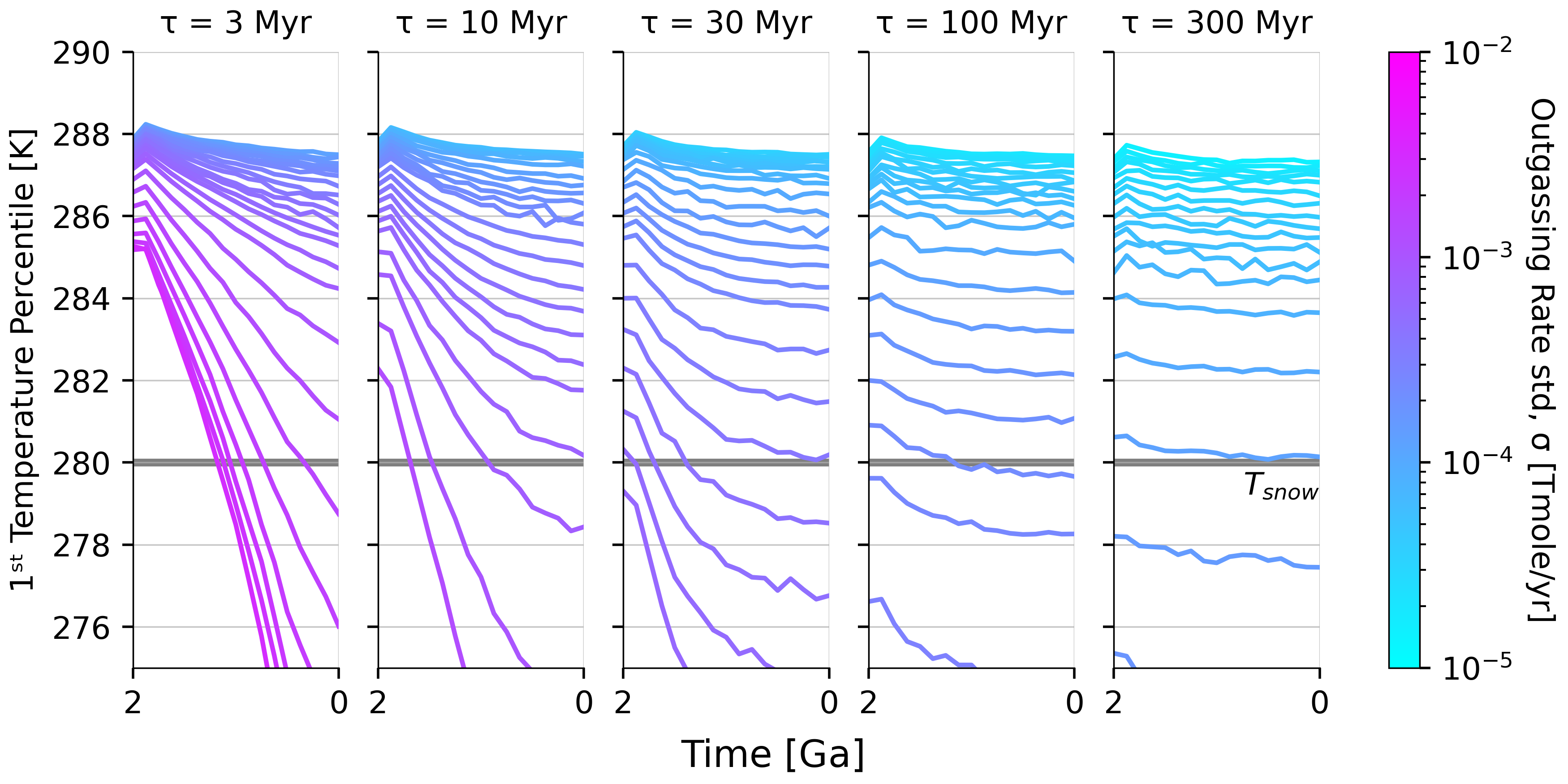}
\caption{The value of the first temperature percentile over time in our ensemble. Moving to the right, each panel shows results for a higher/longer relaxation timescale $\tau$. Within each panel, results for different values of $\sigma$ are delineated by color. The hypothetical snowball temperature ${\Tsnow=280}$~K is shown in each panel. When $\tau$ is small, the outgassing rate is characterized by rapid deviations (see Figure \ref{fig:outexamples}) and the likelihood of cold temperatures increases dramatically as time progresses. As $\tau$ becomes larger, outgassing rates vary over longer periods and the likelihood of cold temperatures becomes much more uniform over time. The first percentile is an arbitrary choice, but other low quantiles demonstrate the same basic relationships between $\tau$, $\sigma$, and the likelihood of low temperatures.}
\label{fig:Tquantiles}
\end{figure}

Figure \ref{fig:Tquantiles} shows the principal results of our ensemble. It shows the first temperature percentile, over time, for different values of $\tau$. The first percentile is a somewhat arbitrary choice. However, the same results are evident for whatever small percentile is chosen. Smaller values of $\tau$ produce more rapid fluctuations in the outgassing rate and the likelihood of cold excursions reaching the snowball temperature increases markedly with time. With higher values of $\tau$, the outgassing history transitions to long-term variability, flattening the first temperature percentile curves.  When $\tau$ is small, snowballs are considerably more likely at later times. When $\tau$ is large, the outgassing rate changes more slowly and snowball likelihood is more uniform over time.

Comparing the leftmost panel of Figure \ref{fig:Tquantiles} to Figure \ref{fig:statprob} is revealing. Low values of $\tau$ in Figure \ref{fig:Tquantiles} recover similar results to \citeA{wordsworth_how_2021}, who found that the likelihood of snowball events dramatically increases over time along with the secular brightening of the Sun (as explained in Section \ref{sec:stochastic_co2}). For larger values of $\tau$, which permit longer-duration excursions in $V$ (Figure \ref{fig:outexamples}), the variation of the 1st percentile temperature over time is much smaller. This is more consistent with the Earth's geological record of snowball episodes, which appear around 2.5 Ga, 720 Ma, and 560 Ma.

The results of this simple model suggest some constraints on the variability of volcanic outgassing over Earth's history. For temperatures to generally remain above $\sim$280 K, yet not to preclude early snowball events, $\sigma$ may have been fairly small with a larger value of $\tau$. Qualitatively speaking, variability in the carbon cycle and atmospheric \cotwo must have considerable long-term drift, not simply short-term excursions. This is because a model with only short-term variability exhibits a very dramatic preference for snowball events later in Earth's history, contradicting the geologic record.

Although it is difficult to reconstruct the history of volcanic outgassing, geological evidence indicates values of $\tau > 100$ Ma may be consistent with the proxy record. \citeA{McKenzie-Horton-Loomis-et-al-2016:continental} used detrital zircon ages as a proxy for continental arc-volcanism over the past 720 Myr and found fluctuations in arc activity over the timescale of hundreds of Myr. Furthermore, studies reconstructing paleogeographic volcanic arc distributions show that variations in arc length of up to a factor of two over may be responsible for icehouse–greenhouse transitions over the past billion years \cite<e.g.,>{Lee-Shen-Slotnick-et-al-2013:continental,Mills-Scotese-Walding-et-al-2017:elevated}. These long-term variations in outgassing rate require $\tau > 100$ Ma to be realized in our stochastic model. 




\section{Conclusion}
\label{sec:conclusion}

In this paper, we discuss simple models of carbon variability deep into Earth's history. One model treats the atmospheric \cotwo concentration directly as a stochastic variable. While instructive, this model is limited in its flexibility and can only produce results that are not compatible with the Earth's history of snowball events. Another model, introduced here, treats carbon sources and sinks independently, and considers the effect of randomly varying \cotwo outgassing with deterministic weathering. This model is slightly more complex, allowing for a range of different behavior that can be compared with the first model and with the snowball record. When outgassing varies over longer periods, specifically with a relaxation time scale of roughly $\tau>100$~Myr, snowballs are possible early in Earth history, which is more compatible with observations.

We note here have been several suggestions regarding the absence of snowball events during the Phanerozoic, including the evolution of land plants, which have been suggested to bring down atmospheric CO$_2$ while simultaneously increasing the strength of the silicate weathering thermostat \cite<e.g.,>{Berner-1994:geocarb, pierrehumbert_principles_2010, Dahl-Arens-2020:impacts}, Another suggestion involves the evolution of fungi and lichens, which are also expected to increase weathering rates \cite{Evans-2003:fundamental,Knauth-Kennedy-2009:late}. The increasing strength of biological regulatory feedbacks may be have been responsible for the change in character of Earth's glacial cycles (from global-scale events to frequent glaciations that do not extend to the low latitudes). Yet, at face value, the model of \citeA{wordsworth_how_2021} predicts that snowball episodes will become increasingly likely in Earth's future with increasing solar luminosity unless stabilizing feedbacks, whether biological or otherwise, strengthen accordingly.

A recent study by \citeA{arnscheidt_routes_2020} discussed different mechanisms for transition into snowball, including fluctuations in radiative forcing and changes in volcanic outgassing or weathering rate. They distinguish between two routes to global glaciation: ``route A'' involving the climate system entering a limit cycle regime, and ``route B'' involving a rapid fluctuation in solar luminosity leading to a solitary ``transient'' glaciation. The negative temperature excursions caused by variable volcanic outgassing in our model can only explain ``route A'' glaciations, and are more consistent with the idea that the Sturtian, Marinoan, and Gaskiers glaciations represent three iterations of a limit cycle \cite<as suggested by>{mills_timing_2011}.

Variability in the carbon cycle has often been characterized as either stochastic \emph{or} deterministic \cite{Westerhold-Marwan-Drury-et-al-2020:astronomically,Arnscheidt-Rothman-2021:asymmetry}. The usual conceptual division is between rapid, unpredictable volcanic forcing (stochastic) and a slow change in the background ``weatherability" of the continental land masses. Although this is sometimes a useful construct, it is important to remember that both of these processes can, and probably do, influence the climate simultaneously. For example, a snowball may be triggered by the combination of slowly increasing weatherability and a relatively sudden change to volcanic outgassing. Our model results with long $\tau$ represent a simple demonstration of this scenario, where outgassing undergoes both long-term drift and occasional shorter-term shocks.

An important caveat of our new model is that we have assumed stochastically varying volcanic outgassing to be the only driver of long-term climate variability. In reality, there are many potential triggers for Snowball events and long-term climate change that are unrelated to outgassing, including changing continental configuration and composition \cite{donnadieu_snowball_2004,cox_continental_2016}, evolution of the biosphere \cite{Hedges-2004:molecular,tziperman_biologically_2011}, sudden radiative forcing from volcanic aerosols \cite{Macdonald-Wordsworth-2017:initiation}, or the collapse of a methane greenhouse \cite{schrag_initiation_2002,Kopp-Kirschvink-Hilburn-et-al-2005:paleoproterozoic}. Indeed, if the Paleoproterozoic event was a transient glaciation caused by aerosol forcing, the timing of the Cryogenian Snowballs late in Earth history would be more consistent with the model of \citeA{wordsworth_how_2021}.

We note that there is a wide range of opportunities to explore randomness in simple climate models. Our model has only considered one form of stochastic outgassing with a deterministic weathering mechanism that is dependent exclusively on temperature. This is, of course, an oversimplification \cite{macdonald_arc-continent_2019, park2021evaluating, baum2022sensitive} and future work could investigate model equations that incorporate variability in other terms in Eq.~\ref{eq:T} such as the planetary albedo and weatherability (e.g., to represent the emplacement of highly weatherable materials and changing continental configuration). For example, it would be straightforward to include a randomly varying weatherability factor that is also represented by an Ornstein-Uhlenbeck process, but this term need not be stochastic at all.

\section{Open Research}
All of the code developed for this project was written in the Julia language \cite{Bezanson2012, bezanson2017julia} and is freely available on GitHub at \url{github.com/markmbaum/random-volcanic-climate} and publicly archived on Zenodo \cite{mark_baum_2022_6799309}. Additional code for the WHAK weathering equation can be found in the \texttt{GEOCLIM.jl} module \cite{mark_baum_2022_6533814}. Plots were created using the open source Matplotlib library \cite{Hunter:2007}.



%
%


\acknowledgments
The computations in this paper were run on the FASRC Cannon cluster supported by the FAS Division of Science Research Computing Group at Harvard University. We thank Kaitlyn Loftus, Camille Hankel, Ned Kleiner, Stephen Bourguet, and Katherine Keller for helpful and collaborative discussions. We thank Dorian Abbot and an anonymous reviewer for helpful feedback that improved the manuscript. 



%
%



\bibliography{manuscript}

\begin{thebibliography}{}

\bibitem [\protect \citeauthoryear {%
Abbot%
}{%
Abbot%
}{%
{\protect \APACyear {2016}}%
}]{%
abbot2016analytical}
\APACinsertmetastar {%
abbot2016analytical}%
\begin{APACrefauthors}%
Abbot, D\BPBI S.%
\end{APACrefauthors}%
\unskip\
\newblock
\APACrefYearMonthDay{2016}{}{}.
\newblock
{\BBOQ}\APACrefatitle {Analytical investigation of the decrease in the size of
  the habitable zone due to a limited CO2 outgassing rate} {Analytical
  investigation of the decrease in the size of the habitable zone due to a
  limited co2 outgassing rate}.{\BBCQ}
\newblock
\APACjournalVolNumPages{The Astrophysical Journal}{827}{2}{117}.
\PrintBackRefs{\CurrentBib}

\bibitem [\protect \citeauthoryear {%
Arnscheidt%
\ \BBA {} Rothman%
}{%
Arnscheidt%
\ \BBA {} Rothman%
}{%
{\protect \APACyear {2020}}%
}]{%
arnscheidt_routes_2020}
\APACinsertmetastar {%
arnscheidt_routes_2020}%
\begin{APACrefauthors}%
Arnscheidt, C\BPBI W.%
\BCBT {}\ \BBA {} Rothman, D\BPBI H.%
\end{APACrefauthors}%
\unskip\
\newblock
\APACrefYearMonthDay{2020}{}{}.
\newblock
{\BBOQ}\APACrefatitle {Routes to global glaciation} {Routes to global
  glaciation}.{\BBCQ}
\newblock
\APACjournalVolNumPages{Proceedings of the Royal Society A: Mathematical,
  Physical and Engineering Sciences}{476}{2239}{}.
\newblock
\begin{APACrefDOI} \doi{10.1098/rspa.2020.0303} \end{APACrefDOI}
\PrintBackRefs{\CurrentBib}

\bibitem [\protect \citeauthoryear {%
Arnscheidt%
\ \BBA {} Rothman%
}{%
Arnscheidt%
\ \BBA {} Rothman%
}{%
{\protect \APACyear {2021}}%
}]{%
Arnscheidt-Rothman-2021:asymmetry}
\APACinsertmetastar {%
Arnscheidt-Rothman-2021:asymmetry}%
\begin{APACrefauthors}%
Arnscheidt, C\BPBI W.%
\BCBT {}\ \BBA {} Rothman, D\BPBI H.%
\end{APACrefauthors}%
\unskip\
\newblock
\APACrefYearMonthDay{2021}{}{}.
\newblock
{\BBOQ}\APACrefatitle {Asymmetry of extreme Cenozoic climate--carbon cycle
  events} {Asymmetry of extreme cenozoic climate--carbon cycle events}.{\BBCQ}
\newblock
\APACjournalVolNumPages{Science Advances}{7}{33}{eabg6864}.
\PrintBackRefs{\CurrentBib}

\bibitem [\protect \citeauthoryear {%
Baum%
}{%
Baum%
}{%
{\protect \APACyear {2022}}%
}]{%
mark_baum_2022_6799309}
\APACinsertmetastar {%
mark_baum_2022_6799309}%
\begin{APACrefauthors}%
Baum, M.%
\end{APACrefauthors}%
\unskip\
\newblock
\APACrefYearMonthDay{2022}{{\APACmonth{07}}}{}.
\newblock
\APACrefbtitle {markmbaum/random-volcanic-climate: release-2.}
  {markmbaum/random-volcanic-climate: release-2.}
\newblock
\APACaddressPublisher{}{Zenodo}.
\newblock
\begin{APACrefURL} \url{https://doi.org/10.5281/zenodo.6799309}
  \end{APACrefURL}
\newblock
\begin{APACrefDOI} \doi{10.5281/zenodo.6799309} \end{APACrefDOI}
\PrintBackRefs{\CurrentBib}

\bibitem [\protect \citeauthoryear {%
Baum%
\ \BBA {} Fu%
}{%
Baum%
\ \BBA {} Fu%
}{%
{\protect \APACyear {2022}}%
}]{%
mark_baum_2022_6533814}
\APACinsertmetastar {%
mark_baum_2022_6533814}%
\begin{APACrefauthors}%
Baum, M.%
\BCBT {}\ \BBA {} Fu, M.%
\end{APACrefauthors}%
\unskip\
\newblock
\APACrefYearMonthDay{2022}{{\APACmonth{05}}}{}.
\newblock
\APACrefbtitle {markmbaum/GEOCLIM.jl: v0.1.12.} {markmbaum/geoclim.jl:
  v0.1.12.}
\newblock
\APACaddressPublisher{}{Zenodo}.
\newblock
\begin{APACrefURL} \url{https://doi.org/10.5281/zenodo.6533814}
  \end{APACrefURL}
\newblock
\begin{APACrefDOI} \doi{10.5281/zenodo.6533814} \end{APACrefDOI}
\PrintBackRefs{\CurrentBib}

\bibitem [\protect \citeauthoryear {%
Baum%
, Fu%
\BCBL {}\ \BBA {} Bourguet%
}{%
Baum%
\ \protect \BOthers {.}}{%
{\protect \APACyear {2022}}%
}]{%
baum2022sensitive}
\APACinsertmetastar {%
baum2022sensitive}%
\begin{APACrefauthors}%
Baum, M.%
, Fu, M.%
\BCBL {}\ \BBA {} Bourguet, S.%
\end{APACrefauthors}%
\unskip\
\newblock
\APACrefYearMonthDay{2022}{}{}.
\newblock
{\BBOQ}\APACrefatitle {Sensitive Dependence of Global Climate to Continental
  Geometry} {Sensitive dependence of global climate to continental
  geometry}.{\BBCQ}
\newblock
\APACjournalVolNumPages{Geophysical Research Letters}{}{}{e2022GL098843}.
\PrintBackRefs{\CurrentBib}

\bibitem [\protect \citeauthoryear {%
Berner%
}{%
Berner%
}{%
{\protect \APACyear {1990}}%
}]{%
Berner-1990:atmospheric}
\APACinsertmetastar {%
Berner-1990:atmospheric}%
\begin{APACrefauthors}%
Berner, R\BPBI A.%
\end{APACrefauthors}%
\unskip\
\newblock
\APACrefYearMonthDay{1990}{}{}.
\newblock
{\BBOQ}\APACrefatitle {Atmospheric carbon dioxide levels over {Phanerozoic}
  time} {Atmospheric carbon dioxide levels over {Phanerozoic} time}.{\BBCQ}
\newblock
\APACjournalVolNumPages{Science}{249}{4975}{1382--1386}.
\PrintBackRefs{\CurrentBib}

\bibitem [\protect \citeauthoryear {%
Berner%
}{%
Berner%
}{%
{\protect \APACyear {1994}}%
}]{%
Berner-1994:geocarb}
\APACinsertmetastar {%
Berner-1994:geocarb}%
\begin{APACrefauthors}%
Berner, R\BPBI A.%
\end{APACrefauthors}%
\unskip\
\newblock
\APACrefYearMonthDay{1994}{}{}.
\newblock
{\BBOQ}\APACrefatitle {GEOCARB II: A revised model of atmospheric CO [sub 2]
  over phanerozoic time} {Geocarb ii: A revised model of atmospheric co [sub 2]
  over phanerozoic time}.{\BBCQ}
\newblock
\APACjournalVolNumPages{American Journal of Science;(United States)}{294}{1}{}.
\PrintBackRefs{\CurrentBib}

\bibitem [\protect \citeauthoryear {%
Berner%
\ \BBA {} Lasaga%
}{%
Berner%
\ \BBA {} Lasaga%
}{%
{\protect \APACyear {1989}}%
}]{%
Berner-Lasaga-1989:modeling}
\APACinsertmetastar {%
Berner-Lasaga-1989:modeling}%
\begin{APACrefauthors}%
Berner, R\BPBI A.%
\BCBT {}\ \BBA {} Lasaga, A\BPBI C.%
\end{APACrefauthors}%
\unskip\
\newblock
\APACrefYearMonthDay{1989}{03}{}.
\newblock
{\BBOQ}\APACrefatitle {Modeling the Geochemical Carbon Cycle} {Modeling the
  geochemical carbon cycle}.{\BBCQ}
\newblock
\APACjournalVolNumPages{Scientific American}{222}{}{74-82}.
\newblock
\begin{APACrefDOI} \doi{10.1038/scientificamerican0389-74} \end{APACrefDOI}
\PrintBackRefs{\CurrentBib}

\bibitem [\protect \citeauthoryear {%
Berner%
, Lasaga%
\BCBL {}\ \BBA {} Garrels%
}{%
Berner%
\ \protect \BOthers {.}}{%
{\protect \APACyear {1983}}%
}]{%
Berner-Lasaga-Garrels-1983:carbonate}
\APACinsertmetastar {%
Berner-Lasaga-Garrels-1983:carbonate}%
\begin{APACrefauthors}%
Berner, R\BPBI A.%
, Lasaga, A\BPBI C.%
\BCBL {}\ \BBA {} Garrels, R\BPBI M.%
\end{APACrefauthors}%
\unskip\
\newblock
\APACrefYearMonthDay{1983}{}{}.
\newblock
{\BBOQ}\APACrefatitle {The carbonate-silicate geochemical cycle and its effect
  on atmospheric carbon dioxide over the past 100 million years} {The
  carbonate-silicate geochemical cycle and its effect on atmospheric carbon
  dioxide over the past 100 million years}.{\BBCQ}
\newblock
\APACjournalVolNumPages{American Journal of Science}{283}{7}{641--683}.
\newblock
\begin{APACrefDOI} \doi{10.2475/ajs.283.7.641} \end{APACrefDOI}
\PrintBackRefs{\CurrentBib}

\bibitem [\protect \citeauthoryear {%
Bezanson%
, Edelman%
, Karpinski%
\BCBL {}\ \BBA {} Shah%
}{%
Bezanson%
\ \protect \BOthers {.}}{%
{\protect \APACyear {2017}}%
}]{%
bezanson2017julia}
\APACinsertmetastar {%
bezanson2017julia}%
\begin{APACrefauthors}%
Bezanson, J.%
, Edelman, A.%
, Karpinski, S.%
\BCBL {}\ \BBA {} Shah, V\BPBI B.%
\end{APACrefauthors}%
\unskip\
\newblock
\APACrefYearMonthDay{2017}{}{}.
\newblock
{\BBOQ}\APACrefatitle {Julia: {A} fresh approach to numerical computing}
  {Julia: {A} fresh approach to numerical computing}.{\BBCQ}
\newblock
\APACjournalVolNumPages{SIAM review}{59}{1}{65--98}.
\newblock
\begin{APACrefURL} \url{https://doi.org/10.1137/141000671} \end{APACrefURL}
\PrintBackRefs{\CurrentBib}

\bibitem [\protect \citeauthoryear {%
Bezanson%
, Karpinski%
, Shah%
\BCBL {}\ \BBA {} Edelman%
}{%
Bezanson%
\ \protect \BOthers {.}}{%
{\protect \APACyear {2012}}%
}]{%
Bezanson2012}
\APACinsertmetastar {%
Bezanson2012}%
\begin{APACrefauthors}%
Bezanson, J.%
, Karpinski, S.%
, Shah, V\BPBI B.%
\BCBL {}\ \BBA {} Edelman, A.%
\end{APACrefauthors}%
\unskip\
\newblock
\APACrefYearMonthDay{2012}{}{}.
\newblock
{\BBOQ}\APACrefatitle {Julia: {A} Fast Dynamic Language for Technical
  Computing} {Julia: {A} fast dynamic language for technical computing}.{\BBCQ}
\newblock
\APACjournalVolNumPages{CoRR}{abs/1209.5145}{}{}.
\newblock
\begin{APACrefURL} \url{http://arxiv.org/abs/1209.5145} \end{APACrefURL}
\PrintBackRefs{\CurrentBib}

\bibitem [\protect \citeauthoryear {%
Cox%
\ \protect \BOthers {.}}{%
Cox%
\ \protect \BOthers {.}}{%
{\protect \APACyear {2016}}%
}]{%
cox_continental_2016}
\APACinsertmetastar {%
cox_continental_2016}%
\begin{APACrefauthors}%
Cox, G\BPBI M.%
, Halverson, G\BPBI P.%
, Stevenson, R\BPBI K.%
, Vokaty, M.%
, Poirier, A.%
, Kunzmann, M.%
\BDBL {}Macdonald, F\BPBI A.%
\end{APACrefauthors}%
\unskip\
\newblock
\APACrefYearMonthDay{2016}{}{}.
\newblock
{\BBOQ}\APACrefatitle {Continental flood basalt weathering as a trigger for
  {Neoproterozoic} {Snowball} {Earth}} {Continental flood basalt weathering as
  a trigger for {Neoproterozoic} {Snowball} {Earth}}.{\BBCQ}
\newblock
\APACjournalVolNumPages{Earth and Planetary Science Letters}{446}{}{}.
\newblock
\begin{APACrefDOI} \doi{10.1016/j.epsl.2016.04.016} \end{APACrefDOI}
\PrintBackRefs{\CurrentBib}

\bibitem [\protect \citeauthoryear {%
Dahl%
\ \BBA {} Arens%
}{%
Dahl%
\ \BBA {} Arens%
}{%
{\protect \APACyear {2020}}%
}]{%
Dahl-Arens-2020:impacts}
\APACinsertmetastar {%
Dahl-Arens-2020:impacts}%
\begin{APACrefauthors}%
Dahl, T\BPBI W.%
\BCBT {}\ \BBA {} Arens, S\BPBI K.%
\end{APACrefauthors}%
\unskip\
\newblock
\APACrefYearMonthDay{2020}{}{}.
\newblock
{\BBOQ}\APACrefatitle {The impacts of land plant evolution on Earth's climate
  and oxygenation state--an interdisciplinary review} {The impacts of land
  plant evolution on earth's climate and oxygenation state--an
  interdisciplinary review}.{\BBCQ}
\newblock
\APACjournalVolNumPages{Chemical Geology}{547}{}{119665}.
\PrintBackRefs{\CurrentBib}

\bibitem [\protect \citeauthoryear {%
Dobrow%
}{%
Dobrow%
}{%
{\protect \APACyear {2016}}%
}]{%
dobrow_introduction_2016}
\APACinsertmetastar {%
dobrow_introduction_2016}%
\begin{APACrefauthors}%
Dobrow, R\BPBI P.%
\end{APACrefauthors}%
\unskip\
\newblock
\APACrefYear{2016}.
\newblock
\APACrefbtitle {Introduction to stochastic processes with {R}} {Introduction to
  stochastic processes with {R}}.
\newblock
\APACaddressPublisher{}{Wiley}.
\PrintBackRefs{\CurrentBib}

\bibitem [\protect \citeauthoryear {%
Donnadieu%
, Goddéris%
, Ramstein%
, Nédélec%
\BCBL {}\ \BBA {} Meert%
}{%
Donnadieu%
\ \protect \BOthers {.}}{%
{\protect \APACyear {2004}}%
}]{%
donnadieu_snowball_2004}
\APACinsertmetastar {%
donnadieu_snowball_2004}%
\begin{APACrefauthors}%
Donnadieu, Y.%
, Goddéris, Y.%
, Ramstein, G.%
, Nédélec, A.%
\BCBL {}\ \BBA {} Meert, J.%
\end{APACrefauthors}%
\unskip\
\newblock
\APACrefYearMonthDay{2004}{}{}.
\newblock
{\BBOQ}\APACrefatitle {A ‘snowball {Earth}’ climate triggered by
  continental break-up through changes in runoff} {A ‘snowball {Earth}’
  climate triggered by continental break-up through changes in runoff}.{\BBCQ}
\newblock
\APACjournalVolNumPages{Nature}{428}{6980}{}.
\newblock
\begin{APACrefDOI} \doi{10.1038/nature02408} \end{APACrefDOI}
\PrintBackRefs{\CurrentBib}

\bibitem [\protect \citeauthoryear {%
Evans%
}{%
Evans%
}{%
{\protect \APACyear {2003}}%
}]{%
Evans-2003:fundamental}
\APACinsertmetastar {%
Evans-2003:fundamental}%
\begin{APACrefauthors}%
Evans, D\BPBI A.%
\end{APACrefauthors}%
\unskip\
\newblock
\APACrefYearMonthDay{2003}{}{}.
\newblock
{\BBOQ}\APACrefatitle {A fundamental Precambrian--Phanerozoic shift in earth's
  glacial style?} {A fundamental precambrian--phanerozoic shift in earth's
  glacial style?}{\BBCQ}
\newblock
\APACjournalVolNumPages{Tectonophysics}{375}{1-4}{353--385}.
\PrintBackRefs{\CurrentBib}

\bibitem [\protect \citeauthoryear {%
Evans%
, Beukes%
\BCBL {}\ \BBA {} Kirschvink%
}{%
Evans%
\ \protect \BOthers {.}}{%
{\protect \APACyear {1997}}%
}]{%
evans_low-latitude_1997}
\APACinsertmetastar {%
evans_low-latitude_1997}%
\begin{APACrefauthors}%
Evans, D\BPBI A.%
, Beukes, N\BPBI J.%
\BCBL {}\ \BBA {} Kirschvink, J\BPBI L.%
\end{APACrefauthors}%
\unskip\
\newblock
\APACrefYearMonthDay{1997}{}{}.
\newblock
{\BBOQ}\APACrefatitle {Low-latitude glaciation in the {Palaeoproterozoic} era}
  {Low-latitude glaciation in the {Palaeoproterozoic} era}.{\BBCQ}
\newblock
\APACjournalVolNumPages{Nature}{386}{6622}{}.
\newblock
\begin{APACrefDOI} \doi{10.1038/386262a0} \end{APACrefDOI}
\PrintBackRefs{\CurrentBib}

\bibitem [\protect \citeauthoryear {%
Farrell%
\ \BBA {} Abbot%
}{%
Farrell%
\ \BBA {} Abbot%
}{%
{\protect \APACyear {2012}}%
}]{%
Farrell-Abbot-2012:mechanism}
\APACinsertmetastar {%
Farrell-Abbot-2012:mechanism}%
\begin{APACrefauthors}%
Farrell, B.%
\BCBT {}\ \BBA {} Abbot, D.%
\end{APACrefauthors}%
\unskip\
\newblock
\APACrefYearMonthDay{2012}{}{}.
\newblock
{\BBOQ}\APACrefatitle {A mechanism for dust-induced destabilization of glacial
  climates} {A mechanism for dust-induced destabilization of glacial
  climates}.{\BBCQ}
\newblock
\APACjournalVolNumPages{Climate of the Past}{8}{6}{2061--2067}.
\PrintBackRefs{\CurrentBib}

\bibitem [\protect \citeauthoryear {%
Feulner%
}{%
Feulner%
}{%
{\protect \APACyear {2012}}%
}]{%
feulner2012faint}
\APACinsertmetastar {%
feulner2012faint}%
\begin{APACrefauthors}%
Feulner, G.%
\end{APACrefauthors}%
\unskip\
\newblock
\APACrefYearMonthDay{2012}{}{}.
\newblock
{\BBOQ}\APACrefatitle {The faint young Sun problem} {The faint young sun
  problem}.{\BBCQ}
\newblock
\APACjournalVolNumPages{Reviews of Geophysics}{50}{2}{}.
\PrintBackRefs{\CurrentBib}

\bibitem [\protect \citeauthoryear {%
Franks%
\ \protect \BOthers {.}}{%
Franks%
\ \protect \BOthers {.}}{%
{\protect \APACyear {2014}}%
}]{%
franks2014new}
\APACinsertmetastar {%
franks2014new}%
\begin{APACrefauthors}%
Franks, P\BPBI J.%
, Royer, D\BPBI L.%
, Beerling, D\BPBI J.%
, Van~de Water, P\BPBI K.%
, Cantrill, D\BPBI J.%
, Barbour, M\BPBI M.%
\BCBL {}\ \BBA {} Berry, J\BPBI A.%
\end{APACrefauthors}%
\unskip\
\newblock
\APACrefYearMonthDay{2014}{}{}.
\newblock
{\BBOQ}\APACrefatitle {New constraints on atmospheric CO2 concentration for the
  Phanerozoic} {New constraints on atmospheric co2 concentration for the
  phanerozoic}.{\BBCQ}
\newblock
\APACjournalVolNumPages{Geophysical Research Letters}{41}{13}{4685--4694}.
\PrintBackRefs{\CurrentBib}

\bibitem [\protect \citeauthoryear {%
Franzke%
, O'Kane%
, Berner%
, Williams%
\BCBL {}\ \BBA {} Lucarini%
}{%
Franzke%
\ \protect \BOthers {.}}{%
{\protect \APACyear {2015}}%
}]{%
franzke2015stochastic}
\APACinsertmetastar {%
franzke2015stochastic}%
\begin{APACrefauthors}%
Franzke, C\BPBI L.%
, O'Kane, T\BPBI J.%
, Berner, J.%
, Williams, P\BPBI D.%
\BCBL {}\ \BBA {} Lucarini, V.%
\end{APACrefauthors}%
\unskip\
\newblock
\APACrefYearMonthDay{2015}{}{}.
\newblock
{\BBOQ}\APACrefatitle {Stochastic climate theory and modeling} {Stochastic
  climate theory and modeling}.{\BBCQ}
\newblock
\APACjournalVolNumPages{Wiley Interdisciplinary Reviews: Climate
  Change}{6}{1}{63--78}.
\PrintBackRefs{\CurrentBib}

\bibitem [\protect \citeauthoryear {%
Gough%
}{%
Gough%
}{%
{\protect \APACyear {1981}}%
}]{%
gough_solar_1981}
\APACinsertmetastar {%
gough_solar_1981}%
\begin{APACrefauthors}%
Gough, D\BPBI O.%
\end{APACrefauthors}%
\unskip\
\newblock
\APACrefYearMonthDay{1981}{}{}.
\newblock
{\BBOQ}\APACrefatitle {Solar interior structure and luminosity variations}
  {Solar interior structure and luminosity variations}.{\BBCQ}
\newblock
\APACjournalVolNumPages{Solar Physics}{}{74}{}.
\PrintBackRefs{\CurrentBib}

\bibitem [\protect \citeauthoryear {%
Graham%
}{%
Graham%
}{%
{\protect \APACyear {2021}}%
}]{%
Graham-2021:high}
\APACinsertmetastar {%
Graham-2021:high}%
\begin{APACrefauthors}%
Graham, R\BPBI J.%
\end{APACrefauthors}%
\unskip\
\newblock
\APACrefYearMonthDay{2021}{}{}.
\newblock
{\BBOQ}\APACrefatitle {High pCO2 reduces sensitivity to CO2 perturbations on
  temperate, Earth-like planets throughout most of habitable zone} {High pco2
  reduces sensitivity to co2 perturbations on temperate, earth-like planets
  throughout most of habitable zone}.{\BBCQ}
\newblock
\APACjournalVolNumPages{Astrobiology}{21}{11}{1406--1420}.
\PrintBackRefs{\CurrentBib}

\bibitem [\protect \citeauthoryear {%
Greenwood%
\ \BBA {} Wing%
}{%
Greenwood%
\ \BBA {} Wing%
}{%
{\protect \APACyear {1995}}%
}]{%
Greenwood-Wing-1995:eocene}
\APACinsertmetastar {%
Greenwood-Wing-1995:eocene}%
\begin{APACrefauthors}%
Greenwood, D\BPBI R.%
\BCBT {}\ \BBA {} Wing, S\BPBI L.%
\end{APACrefauthors}%
\unskip\
\newblock
\APACrefYearMonthDay{1995}{}{}.
\newblock
{\BBOQ}\APACrefatitle {{Eocene} continental climates and latitudinal
  temperature gradients} {{Eocene} continental climates and latitudinal
  temperature gradients}.{\BBCQ}
\newblock
\APACjournalVolNumPages{Geology}{23}{11}{1044--1048}.
\PrintBackRefs{\CurrentBib}

\bibitem [\protect \citeauthoryear {%
Halevy%
\ \BBA {} Bachan%
}{%
Halevy%
\ \BBA {} Bachan%
}{%
{\protect \APACyear {2017}}%
}]{%
Halevy-Bachan-2017:geologic}
\APACinsertmetastar {%
Halevy-Bachan-2017:geologic}%
\begin{APACrefauthors}%
Halevy, I.%
\BCBT {}\ \BBA {} Bachan, A.%
\end{APACrefauthors}%
\unskip\
\newblock
\APACrefYearMonthDay{2017}{}{}.
\newblock
{\BBOQ}\APACrefatitle {The geologic history of seawater pH} {The geologic
  history of seawater ph}.{\BBCQ}
\newblock
\APACjournalVolNumPages{Science}{355}{6329}{1069--1071}.
\PrintBackRefs{\CurrentBib}

\bibitem [\protect \citeauthoryear {%
Hasselmann%
}{%
Hasselmann%
}{%
{\protect \APACyear {1976}}%
}]{%
hasselmann_stochastic_1976}
\APACinsertmetastar {%
hasselmann_stochastic_1976}%
\begin{APACrefauthors}%
Hasselmann, K.%
\end{APACrefauthors}%
\unskip\
\newblock
\APACrefYearMonthDay{1976}{}{}.
\newblock
{\BBOQ}\APACrefatitle {Stochastic climate models {Part} {I}. {Theory}}
  {Stochastic climate models {Part} {I}. {Theory}}.{\BBCQ}
\newblock
\APACjournalVolNumPages{Tellus}{28}{6}{}.
\newblock
\begin{APACrefDOI} \doi{10.1111/j.2153-3490.1976.tb00696.x} \end{APACrefDOI}
\PrintBackRefs{\CurrentBib}

\bibitem [\protect \citeauthoryear {%
Hedges%
}{%
Hedges%
}{%
{\protect \APACyear {2004}}%
}]{%
Hedges-2004:molecular}
\APACinsertmetastar {%
Hedges-2004:molecular}%
\begin{APACrefauthors}%
Hedges, S\BPBI B.%
\end{APACrefauthors}%
\unskip\
\newblock
\APACrefYearMonthDay{2004}{}{}.
\newblock
{\BBOQ}\APACrefatitle {Molecular clocks and a biological trigger for
  Neoproterozoic Snowball Earth events and the Cambrian explosion} {Molecular
  clocks and a biological trigger for neoproterozoic snowball earth events and
  the cambrian explosion}.{\BBCQ}
\newblock
\APACjournalVolNumPages{SYSTEMATICS ASSOCIATION SPECIAL VOLUME}{66}{}{27--40}.
\PrintBackRefs{\CurrentBib}

\bibitem [\protect \citeauthoryear {%
Hoffman%
, Kaufman%
, Halverson%
\BCBL {}\ \BBA {} Schrag%
}{%
Hoffman%
\ \protect \BOthers {.}}{%
{\protect \APACyear {1998}}%
}]{%
Hoffman-Kaufman-Halverson-et-al-1998:neoproterozoic}
\APACinsertmetastar {%
Hoffman-Kaufman-Halverson-et-al-1998:neoproterozoic}%
\begin{APACrefauthors}%
Hoffman, P\BPBI F.%
, Kaufman, A\BPBI J.%
, Halverson, G\BPBI P.%
\BCBL {}\ \BBA {} Schrag, D\BPBI P.%
\end{APACrefauthors}%
\unskip\
\newblock
\APACrefYearMonthDay{1998}{aug 28}{}.
\newblock
{\BBOQ}\APACrefatitle {A {Neoproterozoic} snowball {Earth}} {A {Neoproterozoic}
  snowball {Earth}}.{\BBCQ}
\newblock
\APACjournalVolNumPages{Science}{281}{5381}{1342-1346}.
\PrintBackRefs{\CurrentBib}

\bibitem [\protect \citeauthoryear {%
Hoffman%
\ \BBA {} Schrag%
}{%
Hoffman%
\ \BBA {} Schrag%
}{%
{\protect \APACyear {2002}}%
}]{%
hoffman_snowball_2002}
\APACinsertmetastar {%
hoffman_snowball_2002}%
\begin{APACrefauthors}%
Hoffman, P\BPBI F.%
\BCBT {}\ \BBA {} Schrag, D\BPBI P.%
\end{APACrefauthors}%
\unskip\
\newblock
\APACrefYearMonthDay{2002}{}{}.
\newblock
{\BBOQ}\APACrefatitle {The snowball {Earth} hypothesis: testing the limits of
  global change} {The snowball {Earth} hypothesis: testing the limits of global
  change}.{\BBCQ}
\newblock
\APACjournalVolNumPages{Terra Nova}{14}{3}{}.
\newblock
\begin{APACrefDOI} \doi{10.1046/j.1365-3121.2002.00408.x} \end{APACrefDOI}
\PrintBackRefs{\CurrentBib}

\bibitem [\protect \citeauthoryear {%
Hunter%
}{%
Hunter%
}{%
{\protect \APACyear {2007}}%
}]{%
Hunter:2007}
\APACinsertmetastar {%
Hunter:2007}%
\begin{APACrefauthors}%
Hunter, J\BPBI D.%
\end{APACrefauthors}%
\unskip\
\newblock
\APACrefYearMonthDay{2007}{}{}.
\newblock
{\BBOQ}\APACrefatitle {Matplotlib: A 2D graphics environment} {Matplotlib: A 2d
  graphics environment}.{\BBCQ}
\newblock
\APACjournalVolNumPages{Computing in Science \& Engineering}{9}{3}{90--95}.
\newblock
\begin{APACrefDOI} \doi{10.1109/MCSE.2007.55} \end{APACrefDOI}
\PrintBackRefs{\CurrentBib}

\bibitem [\protect \citeauthoryear {%
Imkeller%
\ \BBA {} Monahan%
}{%
Imkeller%
\ \BBA {} Monahan%
}{%
{\protect \APACyear {2002}}%
}]{%
imkeller2002conceptual}
\APACinsertmetastar {%
imkeller2002conceptual}%
\begin{APACrefauthors}%
Imkeller, P.%
\BCBT {}\ \BBA {} Monahan, A\BPBI H.%
\end{APACrefauthors}%
\unskip\
\newblock
\APACrefYearMonthDay{2002}{}{}.
\newblock
{\BBOQ}\APACrefatitle {Conceptual stochastic climate models} {Conceptual
  stochastic climate models}.{\BBCQ}
\newblock
\APACjournalVolNumPages{Stochastics and Dynamics}{2}{03}{311--326}.
\PrintBackRefs{\CurrentBib}

\bibitem [\protect \citeauthoryear {%
Imkeller%
\ \BBA {} Von~Storch%
}{%
Imkeller%
\ \BBA {} Von~Storch%
}{%
{\protect \APACyear {2001}}%
}]{%
imkeller2001stochastic}
\APACinsertmetastar {%
imkeller2001stochastic}%
\begin{APACrefauthors}%
Imkeller, P.%
\BCBT {}\ \BBA {} Von~Storch, J\BHBI S.%
\end{APACrefauthors}%
\unskip\
\newblock
\APACrefYear{2001}.
\newblock
\APACrefbtitle {Stochastic climate models} {Stochastic climate models}\
  (\BVOL~49).
\newblock
\APACaddressPublisher{}{Springer Science \& Business Media}.
\PrintBackRefs{\CurrentBib}

\bibitem [\protect \citeauthoryear {%
Jacobs%
}{%
Jacobs%
}{%
{\protect \APACyear {2010}}%
}]{%
jacobs_stochastic_2010}
\APACinsertmetastar {%
jacobs_stochastic_2010}%
\begin{APACrefauthors}%
Jacobs, K.%
\end{APACrefauthors}%
\unskip\
\newblock
\APACrefYear{2010}.
\newblock
\APACrefbtitle {Stochastic {Processes} for {Physicists}: {Understanding}
  {Noisy} {Systems}} {Stochastic {Processes} for {Physicists}: {Understanding}
  {Noisy} {Systems}}.
\newblock
\APACaddressPublisher{}{Cambridge University Press}.
\newblock
\begin{APACrefDOI} \doi{10.1017/CBO9780511815980} \end{APACrefDOI}
\PrintBackRefs{\CurrentBib}

\bibitem [\protect \citeauthoryear {%
Kirschvink%
\ \protect \BOthers {.}}{%
Kirschvink%
\ \protect \BOthers {.}}{%
{\protect \APACyear {2000}}%
}]{%
kirschvink_paleoproterozoic_2000}
\APACinsertmetastar {%
kirschvink_paleoproterozoic_2000}%
\begin{APACrefauthors}%
Kirschvink, J\BPBI L.%
, Gaidos, E\BPBI J.%
, Bertani, L\BPBI E.%
, Beukes, N\BPBI J.%
, Gutzmer, J.%
, Maepa, L\BPBI N.%
\BCBL {}\ \BBA {} Steinberger, R\BPBI E.%
\end{APACrefauthors}%
\unskip\
\newblock
\APACrefYearMonthDay{2000}{}{}.
\newblock
{\BBOQ}\APACrefatitle {Paleoproterozoic snowball {Earth}: {Extreme} climatic
  and geochemical global change and its biological consequences}
  {Paleoproterozoic snowball {Earth}: {Extreme} climatic and geochemical global
  change and its biological consequences}.{\BBCQ}
\newblock
\APACjournalVolNumPages{Proceedings of the National Academy of
  Sciences}{97}{4}{}.
\newblock
\begin{APACrefDOI} \doi{10.1073/pnas.97.4.1400} \end{APACrefDOI}
\PrintBackRefs{\CurrentBib}

\bibitem [\protect \citeauthoryear {%
Knauth%
\ \BBA {} Kennedy%
}{%
Knauth%
\ \BBA {} Kennedy%
}{%
{\protect \APACyear {2009}}%
}]{%
Knauth-Kennedy-2009:late}
\APACinsertmetastar {%
Knauth-Kennedy-2009:late}%
\begin{APACrefauthors}%
Knauth, L\BPBI P.%
\BCBT {}\ \BBA {} Kennedy, M\BPBI J.%
\end{APACrefauthors}%
\unskip\
\newblock
\APACrefYearMonthDay{2009}{}{}.
\newblock
{\BBOQ}\APACrefatitle {The late Precambrian greening of the Earth} {The late
  precambrian greening of the earth}.{\BBCQ}
\newblock
\APACjournalVolNumPages{Nature}{460}{7256}{728--732}.
\PrintBackRefs{\CurrentBib}

\bibitem [\protect \citeauthoryear {%
Koll%
\ \BBA {} Cronin%
}{%
Koll%
\ \BBA {} Cronin%
}{%
{\protect \APACyear {2018}}%
}]{%
Koll-Cronin-2018:earth}
\APACinsertmetastar {%
Koll-Cronin-2018:earth}%
\begin{APACrefauthors}%
Koll, D\BPBI D.%
\BCBT {}\ \BBA {} Cronin, T\BPBI W.%
\end{APACrefauthors}%
\unskip\
\newblock
\APACrefYearMonthDay{2018}{}{}.
\newblock
{\BBOQ}\APACrefatitle {Earth’s outgoing longwave radiation linear due to H2O
  greenhouse effect} {Earth’s outgoing longwave radiation linear due to h2o
  greenhouse effect}.{\BBCQ}
\newblock
\APACjournalVolNumPages{Proceedings of the National Academy of
  Sciences}{115}{41}{10293--10298}.
\PrintBackRefs{\CurrentBib}

\bibitem [\protect \citeauthoryear {%
Kopp%
, Kirschvink%
, Hilburn%
\BCBL {}\ \BBA {} Nash%
}{%
Kopp%
\ \protect \BOthers {.}}{%
{\protect \APACyear {2005}}%
}]{%
Kopp-Kirschvink-Hilburn-et-al-2005:paleoproterozoic}
\APACinsertmetastar {%
Kopp-Kirschvink-Hilburn-et-al-2005:paleoproterozoic}%
\begin{APACrefauthors}%
Kopp, R\BPBI E.%
, Kirschvink, J\BPBI L.%
, Hilburn, I\BPBI A.%
\BCBL {}\ \BBA {} Nash, C\BPBI Z.%
\end{APACrefauthors}%
\unskip\
\newblock
\APACrefYearMonthDay{2005}{}{}.
\newblock
{\BBOQ}\APACrefatitle {The Paleoproterozoic snowball Earth: a climate disaster
  triggered by the evolution of oxygenic photosynthesis} {The paleoproterozoic
  snowball earth: a climate disaster triggered by the evolution of oxygenic
  photosynthesis}.{\BBCQ}
\newblock
\APACjournalVolNumPages{Proceedings of the National Academy of
  Sciences}{102}{32}{11131--11136}.
\PrintBackRefs{\CurrentBib}

\bibitem [\protect \citeauthoryear {%
Krissansen-Totton%
, Arney%
\BCBL {}\ \BBA {} Catling%
}{%
Krissansen-Totton%
\ \protect \BOthers {.}}{%
{\protect \APACyear {2018}}%
}]{%
Krissansen-Totton-Arney-Catling-2018:constraining}
\APACinsertmetastar {%
Krissansen-Totton-Arney-Catling-2018:constraining}%
\begin{APACrefauthors}%
Krissansen-Totton, J.%
, Arney, G\BPBI N.%
\BCBL {}\ \BBA {} Catling, D\BPBI C.%
\end{APACrefauthors}%
\unskip\
\newblock
\APACrefYearMonthDay{2018}{}{}.
\newblock
{\BBOQ}\APACrefatitle {Constraining the climate and ocean pH of the early Earth
  with a geological carbon cycle model} {Constraining the climate and ocean ph
  of the early earth with a geological carbon cycle model}.{\BBCQ}
\newblock
\APACjournalVolNumPages{Proceedings of the National Academy of
  Sciences}{115}{16}{4105--4110}.
\PrintBackRefs{\CurrentBib}

\bibitem [\protect \citeauthoryear {%
Lee%
\ \protect \BOthers {.}}{%
Lee%
\ \protect \BOthers {.}}{%
{\protect \APACyear {2013}}%
}]{%
Lee-Shen-Slotnick-et-al-2013:continental}
\APACinsertmetastar {%
Lee-Shen-Slotnick-et-al-2013:continental}%
\begin{APACrefauthors}%
Lee, C\BHBI T\BPBI A.%
, Shen, B.%
, Slotnick, B\BPBI S.%
, Liao, K.%
, Dickens, G\BPBI R.%
, Yokoyama, Y.%
\BDBL {}others%
\end{APACrefauthors}%
\unskip\
\newblock
\APACrefYearMonthDay{2013}{}{}.
\newblock
{\BBOQ}\APACrefatitle {Continental arc--island arc fluctuations, growth of
  crustal carbonates, and long-term climate change} {Continental arc--island
  arc fluctuations, growth of crustal carbonates, and long-term climate
  change}.{\BBCQ}
\newblock
\APACjournalVolNumPages{Geosphere}{9}{1}{21--36}.
\PrintBackRefs{\CurrentBib}

\bibitem [\protect \citeauthoryear {%
Lenardic%
, Jellinek%
, Foley%
, O'Neill%
\BCBL {}\ \BBA {} Moore%
}{%
Lenardic%
\ \protect \BOthers {.}}{%
{\protect \APACyear {2016}}%
}]{%
lenardic2016climate}
\APACinsertmetastar {%
lenardic2016climate}%
\begin{APACrefauthors}%
Lenardic, A.%
, Jellinek, A.%
, Foley, B.%
, O'Neill, C.%
\BCBL {}\ \BBA {} Moore, W.%
\end{APACrefauthors}%
\unskip\
\newblock
\APACrefYearMonthDay{2016}{}{}.
\newblock
{\BBOQ}\APACrefatitle {Climate-tectonic coupling: Variations in the mean,
  variations about the mean, and variations in mode} {Climate-tectonic
  coupling: Variations in the mean, variations about the mean, and variations
  in mode}.{\BBCQ}
\newblock
\APACjournalVolNumPages{Journal of Geophysical Research:
  Planets}{121}{10}{1831--1864}.
\PrintBackRefs{\CurrentBib}

\bibitem [\protect \citeauthoryear {%
Macdonald%
, Swanson-Hysell%
, Park%
, Lisiecki%
\BCBL {}\ \BBA {} Jagoutz%
}{%
Macdonald%
\ \protect \BOthers {.}}{%
{\protect \APACyear {2019}}%
}]{%
macdonald_arc-continent_2019}
\APACinsertmetastar {%
macdonald_arc-continent_2019}%
\begin{APACrefauthors}%
Macdonald, F\BPBI A.%
, Swanson-Hysell, N\BPBI L.%
, Park, Y.%
, Lisiecki, L.%
\BCBL {}\ \BBA {} Jagoutz, O.%
\end{APACrefauthors}%
\unskip\
\newblock
\APACrefYearMonthDay{2019}{}{}.
\newblock
{\BBOQ}\APACrefatitle {Arc-continent collisions in the tropics set {Earth}’s
  climate state} {Arc-continent collisions in the tropics set {Earth}’s
  climate state}.{\BBCQ}
\newblock
\APACjournalVolNumPages{Science}{}{}{}.
\newblock
\begin{APACrefDOI} \doi{10.1126/science.aav5300} \end{APACrefDOI}
\PrintBackRefs{\CurrentBib}

\bibitem [\protect \citeauthoryear {%
Macdonald%
\ \BBA {} Wordsworth%
}{%
Macdonald%
\ \BBA {} Wordsworth%
}{%
{\protect \APACyear {2017}}%
}]{%
Macdonald-Wordsworth-2017:initiation}
\APACinsertmetastar {%
Macdonald-Wordsworth-2017:initiation}%
\begin{APACrefauthors}%
Macdonald, F\BPBI A.%
\BCBT {}\ \BBA {} Wordsworth, R.%
\end{APACrefauthors}%
\unskip\
\newblock
\APACrefYearMonthDay{2017}{}{}.
\newblock
{\BBOQ}\APACrefatitle {Initiation of Snowball Earth with volcanic sulfur
  aerosol emissions} {Initiation of snowball earth with volcanic sulfur aerosol
  emissions}.{\BBCQ}
\newblock
\APACjournalVolNumPages{Geophysical Research Letters}{44}{4}{1938--1946}.
\PrintBackRefs{\CurrentBib}

\bibitem [\protect \citeauthoryear {%
Marshall%
, Walker%
\BCBL {}\ \BBA {} Kuhn%
}{%
Marshall%
\ \protect \BOthers {.}}{%
{\protect \APACyear {1988}}%
}]{%
Marshall-Walker-Kuhn-1988:long}
\APACinsertmetastar {%
Marshall-Walker-Kuhn-1988:long}%
\begin{APACrefauthors}%
Marshall, H\BPBI G.%
, Walker, J\BPBI C\BPBI G.%
\BCBL {}\ \BBA {} Kuhn, W\BPBI R.%
\end{APACrefauthors}%
\unskip\
\newblock
\APACrefYearMonthDay{1988}{}{}.
\newblock
{\BBOQ}\APACrefatitle {Long-term climate change and the geochemical cycle of
  carbon} {Long-term climate change and the geochemical cycle of
  carbon}.{\BBCQ}
\newblock
\APACjournalVolNumPages{Journal of Geophysical Research:
  Atmospheres}{93}{D1}{791--801}.
\PrintBackRefs{\CurrentBib}

\bibitem [\protect \citeauthoryear {%
McKenzie%
\ \protect \BOthers {.}}{%
McKenzie%
\ \protect \BOthers {.}}{%
{\protect \APACyear {2016}}%
}]{%
McKenzie-Horton-Loomis-et-al-2016:continental}
\APACinsertmetastar {%
McKenzie-Horton-Loomis-et-al-2016:continental}%
\begin{APACrefauthors}%
McKenzie, N\BPBI R.%
, Horton, B\BPBI K.%
, Loomis, S\BPBI E.%
, Stockli, D\BPBI F.%
, Planavsky, N\BPBI J.%
\BCBL {}\ \BBA {} Lee, C\BHBI T\BPBI A.%
\end{APACrefauthors}%
\unskip\
\newblock
\APACrefYearMonthDay{2016}{}{}.
\newblock
{\BBOQ}\APACrefatitle {Continental arc volcanism as the principal driver of
  icehouse-greenhouse variability} {Continental arc volcanism as the principal
  driver of icehouse-greenhouse variability}.{\BBCQ}
\newblock
\APACjournalVolNumPages{Science}{352}{6284}{444--447}.
\PrintBackRefs{\CurrentBib}

\bibitem [\protect \citeauthoryear {%
Mills%
, Scotese%
, Walding%
, Shields%
\BCBL {}\ \BBA {} Lenton%
}{%
Mills%
\ \protect \BOthers {.}}{%
{\protect \APACyear {2017}}%
}]{%
Mills-Scotese-Walding-et-al-2017:elevated}
\APACinsertmetastar {%
Mills-Scotese-Walding-et-al-2017:elevated}%
\begin{APACrefauthors}%
Mills, B\BPBI J\BPBI W.%
, Scotese, C\BPBI R.%
, Walding, N\BPBI G.%
, Shields, G\BPBI A.%
\BCBL {}\ \BBA {} Lenton, T\BPBI M.%
\end{APACrefauthors}%
\unskip\
\newblock
\APACrefYearMonthDay{2017}{}{}.
\newblock
{\BBOQ}\APACrefatitle {Elevated CO2 degassing rates prevented the return of
  Snowball Earth during the Phanerozoic} {Elevated co2 degassing rates
  prevented the return of snowball earth during the phanerozoic}.{\BBCQ}
\newblock
\APACjournalVolNumPages{Nature communications}{8}{1}{1--7}.
\PrintBackRefs{\CurrentBib}

\bibitem [\protect \citeauthoryear {%
Mills%
, Watson%
, Goldblatt%
, Boyle%
\BCBL {}\ \BBA {} Lenton%
}{%
Mills%
\ \protect \BOthers {.}}{%
{\protect \APACyear {2011}}%
}]{%
mills_timing_2011}
\APACinsertmetastar {%
mills_timing_2011}%
\begin{APACrefauthors}%
Mills, B\BPBI J\BPBI W.%
, Watson, A\BPBI J.%
, Goldblatt, C.%
, Boyle, R.%
\BCBL {}\ \BBA {} Lenton, T\BPBI M.%
\end{APACrefauthors}%
\unskip\
\newblock
\APACrefYearMonthDay{2011}{}{}.
\newblock
{\BBOQ}\APACrefatitle {Timing of {Neoproterozoic} glaciations linked to
  transport-limited global weathering} {Timing of {Neoproterozoic} glaciations
  linked to transport-limited global weathering}.{\BBCQ}
\newblock
\APACjournalVolNumPages{Nature Geoscience}{4}{12}{}.
\newblock
\begin{APACrefDOI} \doi{10.1038/ngeo1305} \end{APACrefDOI}
\PrintBackRefs{\CurrentBib}

\bibitem [\protect \citeauthoryear {%
Monta{\~n}ez%
\ \protect \BOthers {.}}{%
Monta{\~n}ez%
\ \protect \BOthers {.}}{%
{\protect \APACyear {2016}}%
}]{%
montanez2016climate}
\APACinsertmetastar {%
montanez2016climate}%
\begin{APACrefauthors}%
Monta{\~n}ez, I\BPBI P.%
, McElwain, J\BPBI C.%
, Poulsen, C\BPBI J.%
, White, J\BPBI D.%
, DiMichele, W\BPBI A.%
, Wilson, J\BPBI P.%
\BDBL {}Hren, M\BPBI T.%
\end{APACrefauthors}%
\unskip\
\newblock
\APACrefYearMonthDay{2016}{}{}.
\newblock
{\BBOQ}\APACrefatitle {Climate, pCO2 and terrestrial carbon cycle linkages
  during late Palaeozoic glacial--interglacial cycles} {Climate, pco2 and
  terrestrial carbon cycle linkages during late palaeozoic
  glacial--interglacial cycles}.{\BBCQ}
\newblock
\APACjournalVolNumPages{Nature Geoscience}{9}{11}{824--828}.
\PrintBackRefs{\CurrentBib}

\bibitem [\protect \citeauthoryear {%
Myhre%
, Highwood%
, Shine%
\BCBL {}\ \BBA {} Stordal%
}{%
Myhre%
\ \protect \BOthers {.}}{%
{\protect \APACyear {1998}}%
}]{%
myhre1998new}
\APACinsertmetastar {%
myhre1998new}%
\begin{APACrefauthors}%
Myhre, G.%
, Highwood, E\BPBI J.%
, Shine, K\BPBI P.%
\BCBL {}\ \BBA {} Stordal, F.%
\end{APACrefauthors}%
\unskip\
\newblock
\APACrefYearMonthDay{1998}{}{}.
\newblock
{\BBOQ}\APACrefatitle {New estimates of radiative forcing due to well mixed
  greenhouse gases} {New estimates of radiative forcing due to well mixed
  greenhouse gases}.{\BBCQ}
\newblock
\APACjournalVolNumPages{Geophysical research letters}{25}{14}{2715--2718}.
\PrintBackRefs{\CurrentBib}

\bibitem [\protect \citeauthoryear {%
Park%
, Swanson-Hysell%
, Lisiecki%
\BCBL {}\ \BBA {} Macdonald%
}{%
Park%
\ \protect \BOthers {.}}{%
{\protect \APACyear {2021}}%
}]{%
park2021evaluating}
\APACinsertmetastar {%
park2021evaluating}%
\begin{APACrefauthors}%
Park, Y.%
, Swanson-Hysell, N\BPBI L.%
, Lisiecki, L\BPBI E.%
\BCBL {}\ \BBA {} Macdonald, F\BPBI A.%
\end{APACrefauthors}%
\unskip\
\newblock
\APACrefYearMonthDay{2021}{}{}.
\newblock
{\BBOQ}\APACrefatitle {Evaluating the relationship between the area and
  latitude of large igneous provinces and Earth's long-term climate state}
  {Evaluating the relationship between the area and latitude of large igneous
  provinces and earth's long-term climate state}.{\BBCQ}
\newblock
\APACjournalVolNumPages{Large igneous provinces: A driver of global
  environmental and biotic changes}{}{}{153--168}.
\PrintBackRefs{\CurrentBib}

\bibitem [\protect \citeauthoryear {%
Pierrehumbert%
}{%
Pierrehumbert%
}{%
{\protect \APACyear {2010}}%
}]{%
pierrehumbert_principles_2010}
\APACinsertmetastar {%
pierrehumbert_principles_2010}%
\begin{APACrefauthors}%
Pierrehumbert, R\BPBI T.%
\end{APACrefauthors}%
\unskip\
\newblock
\APACrefYear{2010}.
\newblock
\APACrefbtitle {Principles of {Planetary} {Climate}} {Principles of {Planetary}
  {Climate}}\ (\PrintOrdinal{1}\ \BEd).
\newblock
\APACaddressPublisher{}{Cambridge University Press}.
\newblock
\begin{APACrefDOI} \doi{10.1017/CBO9780511780783} \end{APACrefDOI}
\PrintBackRefs{\CurrentBib}

\bibitem [\protect \citeauthoryear {%
Pierrehumbert%
, Abbot%
, Voigt%
\BCBL {}\ \BBA {} Koll%
}{%
Pierrehumbert%
\ \protect \BOthers {.}}{%
{\protect \APACyear {2011}}%
}]{%
pierrehumbert_climate_2011}
\APACinsertmetastar {%
pierrehumbert_climate_2011}%
\begin{APACrefauthors}%
Pierrehumbert, R\BPBI T.%
, Abbot, D\BPBI S.%
, Voigt, A.%
\BCBL {}\ \BBA {} Koll, D.%
\end{APACrefauthors}%
\unskip\
\newblock
\APACrefYearMonthDay{2011}{}{}.
\newblock
{\BBOQ}\APACrefatitle {Climate of the {Neoproterozoic}} {Climate of the
  {Neoproterozoic}}.{\BBCQ}
\newblock
\APACjournalVolNumPages{Annual Review of Earth and Planetary
  Sciences}{39}{1}{}.
\newblock
\begin{APACrefDOI} \doi{10.1146/annurev-earth-040809-152447} \end{APACrefDOI}
\PrintBackRefs{\CurrentBib}

\bibitem [\protect \citeauthoryear {%
Prave%
, Condon%
, Hoffmann%
, Tapster%
\BCBL {}\ \BBA {} Fallick%
}{%
Prave%
\ \protect \BOthers {.}}{%
{\protect \APACyear {2016}}%
}]{%
Prave-Condon-Hoffmann-et-al-2016:duration}
\APACinsertmetastar {%
Prave-Condon-Hoffmann-et-al-2016:duration}%
\begin{APACrefauthors}%
Prave, A\BPBI R.%
, Condon, D\BPBI J.%
, Hoffmann, K\BPBI H.%
, Tapster, S.%
\BCBL {}\ \BBA {} Fallick, A\BPBI E.%
\end{APACrefauthors}%
\unskip\
\newblock
\APACrefYearMonthDay{2016}{}{}.
\newblock
{\BBOQ}\APACrefatitle {Duration and nature of the {end-Cryogenian} ({Marinoan})
  glaciation} {Duration and nature of the {end-Cryogenian} ({Marinoan})
  glaciation}.{\BBCQ}
\newblock
\APACjournalVolNumPages{Geology}{44}{8}{631--634}.
\PrintBackRefs{\CurrentBib}

\bibitem [\protect \citeauthoryear {%
Rooney%
, Strauss%
, Brandon%
\BCBL {}\ \BBA {} Macdonald%
}{%
Rooney%
\ \protect \BOthers {.}}{%
{\protect \APACyear {2015}}%
}]{%
Rooney-Strauss-Brandon-et-al-2015:cryogenian}
\APACinsertmetastar {%
Rooney-Strauss-Brandon-et-al-2015:cryogenian}%
\begin{APACrefauthors}%
Rooney, A\BPBI D.%
, Strauss, J\BPBI V.%
, Brandon, A\BPBI D.%
\BCBL {}\ \BBA {} Macdonald, F\BPBI A.%
\end{APACrefauthors}%
\unskip\
\newblock
\APACrefYearMonthDay{2015}{}{}.
\newblock
{\BBOQ}\APACrefatitle {A {Cryogenian} chronology: {Two} long-lasting
  synchronous {Neoproterozoic} glaciations} {A {Cryogenian} chronology: {Two}
  long-lasting synchronous {Neoproterozoic} glaciations}.{\BBCQ}
\newblock
\APACjournalVolNumPages{Geology}{43}{5}{459--462}.
\PrintBackRefs{\CurrentBib}

\bibitem [\protect \citeauthoryear {%
Schrag%
, Berner%
, Hoffman%
\BCBL {}\ \BBA {} Halverson%
}{%
Schrag%
\ \protect \BOthers {.}}{%
{\protect \APACyear {2002}}%
}]{%
schrag_initiation_2002}
\APACinsertmetastar {%
schrag_initiation_2002}%
\begin{APACrefauthors}%
Schrag, D\BPBI P.%
, Berner, R\BPBI A.%
, Hoffman, P\BPBI F.%
\BCBL {}\ \BBA {} Halverson, G\BPBI P.%
\end{APACrefauthors}%
\unskip\
\newblock
\APACrefYearMonthDay{2002}{}{}.
\newblock
{\BBOQ}\APACrefatitle {On the initiation of a snowball {Earth}: {Initiation} of
  a {Snowball} {Earth}} {On the initiation of a snowball {Earth}: {Initiation}
  of a {Snowball} {Earth}}.{\BBCQ}
\newblock
\APACjournalVolNumPages{Geochemistry, Geophysics, Geosystems}{3}{6}{}.
\newblock
\begin{APACrefDOI} \doi{10.1029/2001GC000219} \end{APACrefDOI}
\PrintBackRefs{\CurrentBib}

\bibitem [\protect \citeauthoryear {%
Tziperman%
, Halevy%
, Johnston%
, Knoll%
\BCBL {}\ \BBA {} Schrag%
}{%
Tziperman%
\ \protect \BOthers {.}}{%
{\protect \APACyear {2011}}%
}]{%
tziperman_biologically_2011}
\APACinsertmetastar {%
tziperman_biologically_2011}%
\begin{APACrefauthors}%
Tziperman, E.%
, Halevy, I.%
, Johnston, D\BPBI T.%
, Knoll, A\BPBI H.%
\BCBL {}\ \BBA {} Schrag, D\BPBI P.%
\end{APACrefauthors}%
\unskip\
\newblock
\APACrefYearMonthDay{2011}{}{}.
\newblock
{\BBOQ}\APACrefatitle {Biologically induced initiation of {Neoproterozoic}
  snowball-{Earth} events} {Biologically induced initiation of {Neoproterozoic}
  snowball-{Earth} events}.{\BBCQ}
\newblock
\APACjournalVolNumPages{Proceedings of the National Academy of
  Sciences}{108}{37}{}.
\newblock
\begin{APACrefDOI} \doi{10.1073/pnas.1016361108} \end{APACrefDOI}
\PrintBackRefs{\CurrentBib}

\bibitem [\protect \citeauthoryear {%
Urey%
}{%
Urey%
}{%
{\protect \APACyear {1952}}%
}]{%
Urey-1952:early}
\APACinsertmetastar {%
Urey-1952:early}%
\begin{APACrefauthors}%
Urey, H\BPBI C.%
\end{APACrefauthors}%
\unskip\
\newblock
\APACrefYearMonthDay{1952}{}{}.
\newblock
{\BBOQ}\APACrefatitle {On the early chemical history of the earth and the
  origin of life} {On the early chemical history of the earth and the origin of
  life}.{\BBCQ}
\newblock
\APACjournalVolNumPages{Proceedings of the National Academy of
  Sciences}{38}{4}{351}.
\PrintBackRefs{\CurrentBib}

\bibitem [\protect \citeauthoryear {%
Walker%
, Hays%
\BCBL {}\ \BBA {} Kasting%
}{%
Walker%
\ \protect \BOthers {.}}{%
{\protect \APACyear {1981}}%
}]{%
Walker-Hays-Kasting-1981:negative}
\APACinsertmetastar {%
Walker-Hays-Kasting-1981:negative}%
\begin{APACrefauthors}%
Walker, J\BPBI C\BPBI G.%
, Hays, P\BPBI B.%
\BCBL {}\ \BBA {} Kasting, J\BPBI F.%
\end{APACrefauthors}%
\unskip\
\newblock
\APACrefYearMonthDay{1981}{}{}.
\newblock
{\BBOQ}\APACrefatitle {A negative feedback mechanism for the long-term
  stabilization of {Earth's} surface temperature} {A negative feedback
  mechanism for the long-term stabilization of {Earth's} surface
  temperature}.{\BBCQ}
\newblock
\APACjournalVolNumPages{Journal of Geophysical Research:
  Oceans}{86}{C10}{9776--9782}.
\PrintBackRefs{\CurrentBib}

\bibitem [\protect \citeauthoryear {%
Westerhold%
\ \protect \BOthers {.}}{%
Westerhold%
\ \protect \BOthers {.}}{%
{\protect \APACyear {2020}}%
}]{%
Westerhold-Marwan-Drury-et-al-2020:astronomically}
\APACinsertmetastar {%
Westerhold-Marwan-Drury-et-al-2020:astronomically}%
\begin{APACrefauthors}%
Westerhold, T.%
, Marwan, N.%
, Drury, A\BPBI J.%
, Liebrand, D.%
, Agnini, C.%
, Anagnostou, E.%
\BDBL {}others%
\end{APACrefauthors}%
\unskip\
\newblock
\APACrefYearMonthDay{2020}{}{}.
\newblock
{\BBOQ}\APACrefatitle {An astronomically dated record of Earth’s climate and
  its predictability over the last 66 million years} {An astronomically dated
  record of earth’s climate and its predictability over the last 66 million
  years}.{\BBCQ}
\newblock
\APACjournalVolNumPages{Science}{369}{6509}{1383--1387}.
\PrintBackRefs{\CurrentBib}

\bibitem [\protect \citeauthoryear {%
Wordsworth%
}{%
Wordsworth%
}{%
{\protect \APACyear {2021}}%
}]{%
wordsworth_how_2021}
\APACinsertmetastar {%
wordsworth_how_2021}%
\begin{APACrefauthors}%
Wordsworth, R.%
\end{APACrefauthors}%
\unskip\
\newblock
\APACrefYearMonthDay{2021}{}{}.
\newblock
{\BBOQ}\APACrefatitle {How likely are {Snowball} episodes near the inner edge
  of the habitable zone?} {How likely are {Snowball} episodes near the inner
  edge of the habitable zone?}{\BBCQ}
\newblock
\APACjournalVolNumPages{The Astrophysical Journal Letters}{912}{1}{}.
\newblock
\begin{APACrefDOI} \doi{10.3847/2041-8213/abf7c7} \end{APACrefDOI}
\PrintBackRefs{\CurrentBib}

\bibitem [\protect \citeauthoryear {%
Zeebe%
, Westerhold%
, Littler%
\BCBL {}\ \BBA {} Zachos%
}{%
Zeebe%
\ \protect \BOthers {.}}{%
{\protect \APACyear {2017}}%
}]{%
Zeebe-Westerhold-Littler-et-al-2017:orbital}
\APACinsertmetastar {%
Zeebe-Westerhold-Littler-et-al-2017:orbital}%
\begin{APACrefauthors}%
Zeebe, R\BPBI E.%
, Westerhold, T.%
, Littler, K.%
\BCBL {}\ \BBA {} Zachos, J\BPBI C.%
\end{APACrefauthors}%
\unskip\
\newblock
\APACrefYearMonthDay{2017}{}{}.
\newblock
{\BBOQ}\APACrefatitle {Orbital forcing of the Paleocene and Eocene carbon
  cycle} {Orbital forcing of the paleocene and eocene carbon cycle}.{\BBCQ}
\newblock
\APACjournalVolNumPages{Paleoceanography}{32}{5}{440--465}.
\PrintBackRefs{\CurrentBib}

\end{thebibliography}

%
%
%
%
%

\end{document}